\documentclass[twocolumn]{aastex631}

\usepackage{amsmath}
\usepackage{lipsum}
\usepackage{hyperref}
\usepackage{xcolor}
\usepackage{soul}

\newcommand{\Rsun}{$\text{R}_\odot$}
\newcommand{\Msun}{$\text{M}_\odot$}
\newcommand{\Rearth}{$\text{R}_\oplus$}
\newcommand{\Mearth}{$\text{M}_\oplus$}
\newcommand{\HIP}{HIP\,41378\,\,f}

\shorttitle{Stability and detectability of exomoons orbiting \HIP}
\shortauthors{Harada et al.}

\graphicspath{{./}{figures/}}

\begin{document}

\title{Stability and detectability of exomoons orbiting \HIP, a temperate Jovian planet with an anomalously low apparent density}

\correspondingauthor{Caleb K. Harada}
\email{charada@berkeley.edu}
 
\author[0000-0001-5737-1687]{Caleb K. Harada}
\altaffiliation{NSF Graduate Research Fellow}
\affiliation{Department of Astronomy, 501 Campbell Hall \#3411, University of California, Berkeley, Berkeley, CA 94720, USA}

\author[0000-0001-8189-0233]{Courtney D. Dressing}
\affiliation{Department of Astronomy, 501 Campbell Hall \#3411, University of California, Berkeley, Berkeley, CA 94720, USA}

\author[0000-0003-4157-832X]{Munazza K. Alam}
\affiliation{Earth \& Planets Laboratory, Carnegie Institution for Science, 5241 Broad Branch Rd NW, Washington, DC 20015, USA}

\author[0000-0002-4207-6615]{James Kirk}
\affiliation{Department of Physics, Imperial College London, Prince Consort Road, London, SW7 2AZ, UK}

\author[0000-0003-3204-8183]{Mercedes L\'opez-Morales} \affiliation{Center for Astrophysics ${\rm \mid}$ Harvard {\rm \&} Smithsonian, 60 Garden St, Cambridge, MA 02138, USA}

\author[0000-0003-3290-6758]{Kazumasa Ohno}
\affiliation{Division of Science, National Astronomical Observatory of Japan, Tokyo, Japan}
\affiliation{Department of Astronomy and Astrophysics, University of California, Santa Cruz, 1156 High Street, Santa Cruz, CA 95064, USA}


\author[0000-0001-6519-1598]{Babatunde Akinsanmi}
\affiliation{Observatoire  Astronomique  de  l'Universit\'{e}  de  Gen\`{e}ve, Chemin Pegasi 51, CH-1290 Versoix, Switzerland}

\author[0000-0003-2434-3625]{Susana C. C. Barros}
\affiliation{Instituto de Astrof\'isica e Ci\^encias do Espa\c{c}o, Universidade do Porto, CAUP, Rua das Estrelas, 4150-762 Porto, Portugal}
\affiliation{Departamento de F\'isica e Astronomia, Faculdade de Ci\^encias, Universidade do Porto, Rua do Campo Alegre, 4169-007 Porto, Portugal}

\author[0000-0003-1605-5666]{Lars A. Buchhave}
\affiliation{DTU Space,  National Space Institute, Technical University of Denmark, Elektrovej 328, DK-2800 Kgs. Lyngby, Denmark}

\author[0000-0002-8863-7828]{A.~Collier Cameron}
\affiliation{Centre for Exoplanet Science, SUPA School of Physics and Astronomy, University of St Andrews, North Haugh, St Andrews KY16 9SS, UK}

\author{Ian J.\ M.\ Crossfield}
\affiliation{Department of Physics and Astronomy, University of  Kansas, Lawrence, KS, USA}

\author[0000-0002-8958-0683]{Fei Dai}
\altaffiliation{NASA Sagan Fellow}
\affiliation{Division of Geological and Planetary Sciences, 1200 E California Blvd, Pasadena, CA, 91125, USA}
\affiliation{Department of Astronomy, California Institute of Technology, Pasadena, CA 91125, USA}

\author[0000-0002-8518-9601]{Peter Gao}
\affiliation{Earth \& Planets Laboratory, Carnegie Institution for Science, 5241 Broad Branch Rd NW, Washington, DC 20015, USA}

\author[0000-0002-8965-3969]{Steven Giacalone}
\affiliation{Department of Astronomy, 501 Campbell Hall \#3411, University of California, Berkeley, Berkeley, CA 94720, USA}

\author[0000-0002-2805-5869]{Salom\'{e} Grouffal}
\affiliation{Aix Marseille University, CNRS, CNES, LAM, Marseille, France}

\author[0000-0003-3742-1987]{Jorge Lillo-Box}
\affiliation{Centro de Astrobiolog\'ia (CAB, CSIC-INTA), Depto. de Astrof\'isica, ESAC campus, 28692, Villanueva de la Ca\~nada (Madrid), Spain}

\author[0000-0002-7216-2135]{Andrew W. Mayo}
\affiliation{Department of Astronomy, 501 Campbell Hall \#3411, University of California, Berkeley, Berkeley, CA 94720, USA}
\affiliation{Centre for Star and Planet Formation, Natural History Museum of Denmark \& Niels Bohr Institute, University of Copenhagen, \O ster Voldgade 5-7, DK-1350 Copenhagen K., Denmark}

\author[0000-0001-7254-4363]{Annelies Mortier}
\affiliation{School of Physics \& Astronomy, University of Birmingham, Edgbaston, Birmingham B15 2TT, UK}

\author[0000-0002-3586-1316]{Alexandre Santerne}
\affiliation{Aix Marseille University, CNRS, CNES, LAM, Marseille, France}

\author[0000-0003-4422-2919]{Nuno C. Santos}
\affiliation{Instituto de Astrof\'isica e Ci\^encias do Espa\c{c}o, Universidade do Porto, CAUP, Rua das Estrelas, 4150-762 Porto, Portugal}
\affiliation{Departamento de F\'isica e Astronomia, Faculdade de Ci\^encias, Universidade do Porto, Rua do Campo Alegre, 4169-007 Porto, Portugal}

\author[0000-0001-9047-2965]{S\'ergio G. Sousa}
\affiliation{Instituto de Astrof\'isica e Ci\^encias do Espa\c{c}o, Universidade do Porto, CAUP, Rua das Estrelas, 4150-762 Porto, Portugal}

\author[0000-0002-1845-2617]{Emma V. Turtelboom}
\affiliation{Department of Astronomy, 501 Campbell Hall \#3411, University of California, Berkeley, Berkeley, CA 94720, USA}

\author[0000-0001-7246-5438]{Andrew Vanderburg}
\affiliation{Department of Physics and Kavli Institute for Astrophysics and Space Research, Massachusetts Institute of Technology, Cambridge, MA 02139, USA}

\author[0000-0003-1452-2240]{Peter J. Wheatley}
\affiliation{Centre for Exoplanets and Habitability, University of Warwick, Gibbet Hill Road, CV4 7AL Coventry, UK}
\affiliation{Department of Physics, University of Warwick, Gibbet Hill Road, CV4 7AL Coventry, UK}

\begin{abstract}
Moons orbiting exoplanets (``exomoons'') may hold clues about planet formation, migration, and habitability. In this work, we investigate the plausibility of exomoons orbiting the temperate ($T_\text{eq}=294$ K) giant ($R = 9.2$ \Rearth) planet \HIP, which has been shown to have a low apparent bulk density of $0.09\,\text{g}\,\text{cm}^{-3}$ and a flat near-infrared transmission spectrum, hinting that it may possess circumplanetary rings. Given this planet's long orbital period ($P\approx1.5$ yr), it has been suggested that it may also host a large exomoon. Here, we analyze the orbital stability of a hypothetical exomoon with a satellite-to-planet mass ratio of 0.0123 orbiting \HIP. Combining a new software package, \texttt{astroQTpy}, with \texttt{REBOUND} and \texttt{EqTide}, we conduct a series of N-body and tidal migration simulations, demonstrating that satellites up to this size are largely stable against dynamical escape and collisions. We simulate the expected transit signal from this hypothetical exomoon and show that current transit observations likely cannot constrain the presence of exomoons orbiting \HIP, though future observations may be capable of detecting exomoons in other systems. Finally, we model the combined transmission spectrum of \HIP~and a hypothetical moon with a low-metallicity atmosphere, and show that the total effective spectrum would be contaminated at the $\sim$10 ppm level. Our work not only demonstrates the feasibility of exomoons orbiting \HIP, but also shows that large exomoons may be a source of uncertainty in future high-precision measurements of exoplanet systems.
\end{abstract}

\keywords{Exoplanet astronomy (486), Exoplanet dynamics (490), Exoplanet systems (484), Exoplanet tides (497), Natural satellites (Extrasolar) (483), Transits (1711), Transmission spectroscopy (2133)}

\received{17 March 2023}
\revised{5 September 2023}
\accepted{5 October 2023}
\submitjournal{The Astronomical Journal}

\section{Introduction} \label{sec:intro}

The discovery and characterization of extrasolar moons (i.e., ``exomoons'') will play an important role in developing a more complete understanding of planetary systems. As remnants of planet formation, moons may provide useful information about the formation and evolution of exoplanets \citep[e.g.,][]{Sasaki+2010, Heller+2014_review, Spalding+2016, Ronnet+2018}. Moon-planet interactions may also be crucial for the long-term stability of exoplanet habitability; for example, moons may stabilize planetary obliquity \citep{Laskar+1993, Lissauer+2012} and drive tidal heating \citep[e.g.,][]{Piro+2018}, thereby significantly affecting planetary climate. Moreover, sufficiently heated exomoons may themselves be capable of hosting life, providing an alternative path to habitability beyond conventional habitable-zone terrestrial planets \citep[e.g.,][]{Heller+Barnes_2013, Heller+2014_review}.

Despite the large number of confirmed exoplanets surveyed to date ($>5,000$), there has yet to be a confirmed detection of an exomoon, largely due to the minuscule signals that putative moons imprint on measurements of their exoplanet hosts. Though tentative evidence for exomoons has recently been claimed \citep{Teachey+2018_nature, Kipping+2022}, robustly detecting exomoons is likely beyond the reach of current technology. Nonetheless, theoretical constraints on exomoon formation and evolution can provide valuable insight into plausible satellite-hosting exoplanets, and aide in the interpretation of future high-precision observations. For example, numerical N-body simulations have set limits on exomoon eccentricities and semi-major axes that are plausibly stable over long timescales \citep[e.g.,][]{Domingos+2006, Payne+2013, Rosario-Franco+2020AJ}. Moreover, previous studies have investigated how tidal interactions between bodies can result in the radial migration of an exomoon, leading to its eventual escape or tidal disruption by the planet \citep[e.g.,][]{Barnes+OBrien_2002, Sasaki+2012, Sasaki+Barnes_2014, Piro+2018, Quarles+2020, Jagtap+2021, Sucerquia+2019MNRAS}, as well as the effects of planetary migration on exomoon stability \citep[e.g.,][]{Alvarado-Montes+2017, Sucerquia+2020MNRAS}.

In this work, we investigate the plausibility of exomoons orbiting the temperate low-mass Jovian planet \HIP. HIP\,41378 is a bright ($V \approx 8.9$), slightly evolved late F-type star \citep[$T_\text{eff}=6320$ K, $\log g = 4.294$, $\text{[Fe/H]} = -0.10$, $M_\star = 1.16$ \Msun;][]{Santerne+2019} hosting five known transiting exoplanets \citep{Vanderburg+2016_hip, Berardo+2019}. The system was initially observed over the course of about 75 days during Campaign 5 (C5) of the extended \textit{Kepler} mission \citep[K2;][]{Borucki+2010, Howell+2014}, after which the two innermost planets were classified as sub-Neptunes ($R_\text{p,b} \approx 2.9$ \Rearth, $R_\text{p,c} \approx 2.6$ \Rearth) orbiting in a near 2:1 resonance at approximately 15.6 and 31.7 days \citep{Vanderburg+2016_hip}. However, only single transits were initially observed for the three outer planets ($R_\text{p,d} \approx 4.0$ \Rearth, $R_\text{p,e} \approx 5.5$ \Rearth, $R_\text{p,f} \approx 10$ \Rearth), and hence their orbital periods and ephemerides could not be well constrained \citep{Vanderburg+2016_hip}.

More precise limits on the orbits of planets d and f were later derived after one additional transit was observed for each planet during C18 of the K2 mission \citep{Becker+2019, Berardo+2019}. A subsequent third transit observation of planet f with the ground-based Next Generation Transit Survey \citep[NGTS;][]{Wheatley+2018MNRAS} confirmed its orbital period and suggested the presence of transit timing variations \citep[TTVs;][]{Bryant+2021}. Furthermore, radial velocity monitoring of the host star over hundreds of epochs was used to constrain the masses of all of the planets in the system, providing new insights into their physical properties. \citet{Santerne+2019} conducted a joint analysis using the two campaigns of K2 photometry together with 464 epochs of radial velocity measurements from the HARPS, HARPS-N, HIRES, and PFS spectrographs, and reported updated planet radii, masses, bulk densities, and orbital parameters for the five transiting planets, as well as a tentative detection of a sixth non-transiting planet. However, we note that an updated analysis of the data likely rules out the hypothesis of a tentative sixth planet in the HIP\,41378 system (A. Santerne, private communication). A summary of the most recent planet parameters from \citet{Santerne+2019} is shown in Table \ref{tab:planets}\footnote{An additional transit of \HIP~was observed in November 2022 by a coordinated ground-based effort.}.

The largest and outermost planet in the system, \HIP, orbits near the inner edge of the system's habitable zone with a period of about $1.5$ years. Mysteriously, the planet has an inferred radius of 9.2 \Rearth~but it has a mass of only 12 \Mearth, making it one of the lowest bulk density planets currently known \citep[$\rho = 0.09 \,\text{g}\,\text{cm}^{-3}$;][]{Santerne+2019}. It has been proposed that such low bulk densities in exoplanets could be explained by atmospheric outflows entraining optically thick photochemical hazes, which act to increase the planet's apparent radius in transit \citep{Ohno+2021, Gao+zhang_2020, Wang+Dai_2019ApJ}. However, this mechanism may not be compatible with \HIP~because of the planet's relatively high mass and cool equilibrium temperature \citep[e.g., see Equation 34 of][]{Ohno+2021}. Moreover, \citet{Belkovski+2022} showed that \HIP~would require a very high envelope mass fraction ($\gtrsim 75\%$) to remain in hydrostatic balance, in tension with the core accretion paradigm of planet formation.

An alternative explanation for \HIP's apparently low density is that the planet itself is significantly smaller, but is accompanied by a system of optically thick circumplanetary rings \citep{Zuluaga+2015ApJ, Akinsanmi+2020, Piro+Vissapragada_2020}. Such a ring system, if viewed at sufficiently high inclination, would effectively increase the observed transit depth while allowing the planet to have a bulk density more typical of sub-Neptunes. For example, \citet{Akinsanmi+2020} demonstrated that observations of \HIP~are consistent with a  3.7 \Rearth~($\rho = 1.2 \,\text{g}\,\text{cm}^{-3}$) planet with opaque rings extending from 1.05 to 2.6 planet radii. In addition, the presence of optically thick rings potentially causes a featureless transmission spectrum \citep{Ohno+2022,Ohno+22b}. Interestingly, our recent observations of \HIP~taken with the Widefield Camera 3 (WFC3) aboard the Hubble Space Telescope (HST) reveal a featureless near-infrared ($1.1-1.7$ $\mu$m) transmission spectrum \citep[HST-GO Program 16267;][]{Alam+2022}. Within the precision of these observations, we currently cannot distinguish between circumplanetary rings ($\chi_r^2 \approx 1.03$), high-altitude photochemical hazes ($\chi_r^2 \approx 0.97$), or a high-metallicity atmosphere ($\chi_r^2 \approx 1.84$). Future spectral observations at mid-infrared wavelengths (e.g., with JWST/MIRI) may be able to distinguish between these scenarios \citep{Alam+2022}.

The possibility of a circumplanetary ring system around \HIP~naturally raises the question of whether the planet could also host moons. For example, \citet{Saillenfest+2023} concluded that the migration of a former moon is a viable formation pathway for the proposed ring and tilt of HIP 41378 f. More generally, moons and circumplanetary rings are ubiquitous in the outer Solar System, and earlier studies have suggested that rings are originally formed from moons \citep[e.g.,][]{Canup+2010}, or vice versa \citep[e.g.,][]{Crida+Sharnoz2012}.

Because the Hill radius scales linearly with the distance of a planet from its host star, \HIP's large separation from its host star relative to the population of hot Jupiters ($P \lesssim 10$ d) makes it a favorable environment for exomoons. While \HIP~itself is likely not habitable due to its high gas fraction, its temperate orbital separation raises the possibility that it hosts habitable exomoons. We note that the habitability of icy moons in the outer Solar System is still an open question. Though the detailed level of characterization required to assess exomoon habitability, or even confirm exomoon detections at all, is beyond current instrumental sensitivity, it may be prudent to consider the extent to which the presence of exomoons could manifest as a source of uncertainty in measurements of cool gaseous exoplanets, especially as we look toward future high-precision facilities such as ground-based ELTs and large space-based UV/optical/IR observatories \citep[e.g.,][]{Dalba+2015, LUVOIR_2019arXiv}.

Here we assess the stability of exomoons orbiting \HIP~using a suite of numerical N-body simulations and models of tidal evolution. Then, from the results of our stability analysis, we simulate the observable impact of a large exomoon on the white light curve and infrared transmission spectrum of \HIP. The rest of this paper is organized as follow: in Section \ref{sec:methods} we describe our theoretical methods used to constrain the allowable range of parameters for a stable moon orbiting \HIP. In Section \ref{sec:results_disc}, we present the results from our analysis and discuss the future observational implications. We conclude with a summary in Section \ref{sec:conclusion}.

\begin{deluxetable*}{lCCC}[thb]
\tablecolumns{4}
\tablefontsize{\footnotesize}
\tablecaption{HIP\,41378 system parameters from \citet{Santerne+2019}. \label{tab:planets}}
\tablehead{
    \colhead{Parameter} &
    \multicolumn{3}{C}{\text{Median and 68.3\% credible interval}}
}
\startdata
\textit{\textbf{Planet parameters}} & \text{\underline{HIP\,41378\,\,b}} & \text{\underline{HIP\,41378\,\,c}} & \text{\underline{HIP\,41378\,\,d}} \\
Orbital period, $P$ (days) & 15.57208 \pm 2 \times 10^{-5} & 31.70603 \pm 6 \times 10^{-5} & 278.3618 \pm 5 \times 10^{-4} \\
Eccentricity, $e$ & 0.07 \pm 0.06 & 0.04^{+0.04}_{-0.03} & 0.06 \pm 0.06 \\
Semi-major axis, $a$ (AU) & 0.1283 \pm 1.5 \times 10^{-3} & 0.2061 \pm 2.4 \times 10^{-3} & 0.88 \pm 0.01 \\
Inclination, $i$ ($^\circ$) & 88.75 \pm 0.13 & 88.477^{+0.035}_{-0.061} & 89.80 \pm 0.02 \\
Radius, $R_p$ (\Rearth) & 2.595 \pm 0.036 & 2.727 \pm 0.060 & 3.54 \pm 0.06 \\
Mass, $M_p$ (\Mearth) & 6.89 \pm 0.88 & 4.4 \pm 1.1 & < 4.6^\dagger \\
Bulk density, $\rho_p$ (g cm$^{-3}$) & 2.17 \pm 0.28 & 1.19 \pm 0.30 & < 0.56^\dagger \\
Equilibrium temperature$^*$, $T_\text{eq}$ (K) & 959^{+9}_{-5} & 757^{+7}_{-4} & 367^{+3}_{-2} \\
Insolation flux, $S$ ($S_\oplus$) & 140^{+5}_{-3} & 54^{+2}_{-1} & 3.01^{+0.11}_{-0.06} \\
& & & \\
\textit{\textbf{Planet parameters}} & \text{\underline{HIP\,41378\,\,e}} & \text{\underline{HIP\,41378\,\,f}} & \\
Orbital period, $P$ (days) & 369 \pm 10 & 542.07975 \pm 1.4 \times 10^{-4} & \\
Eccentricity, $e$ & 0.14 \pm 0.09 & 0.004^{+0.009}_{-0.003} & \\
Semi-major axis, $a$ (AU) & 1.06^{+0.03}_{-0.02} & 1.37 \pm 0.02 & \\
Inclination, $i$ ($^\circ$) & 89.84^{+0.07}_{-0.03} & 89.971^{+0.01}_{-0.008} & \\
Radius, $R_p$ (\Rearth) & 4.92 \pm 0.09 & 9.2 \pm 0.1 & \\
Mass, $M_p$ (\Mearth) & 12 \pm 5 & 12 \pm 3 & \\
Bulk density, $\rho_p$ (g cm$^{-3}$) & 0.55 \pm 0.23 & 0.09 \pm 0.02 & \\
Equilibrium temperature$^*$, $T_\text{eq}$ (K) & 335 \pm 4 & 294^{+3}_{-1} & \\
Insolation flux, $S$ ($S_\oplus$) & 2.1 \pm 0.1 & 1.24^{+0.05}_{-0.02} & \\
& & & \\
\textit{\textbf{Stellar parameters}} & \text{\underline{HIP\,41378}} & & \\
{Effective temperature, $T_\text{eff}$ (K)} & {6320_{-30}^{+60}} & & \\
{Surface gravity, $\log g$ (cgs)} & {4.294 \pm 0.006} & & \\
{Metallicity, [Fe/H] (dex)} & {-0.10 \pm 0.07} & & \\
{Mass, $M_\star$ (\Msun)} & {1.16 \pm 0.04}  & & \\
{Radius, $R_\star$ (\Rsun)} &{1.273 \pm 0.015} & & \\
{Age, $\tau_\star$ (Gyr)} & {3.1 \pm 0.6} & & \\
{Distance from Earth, $D$} (pc) & {103 \pm 2} & & \\
\enddata
\tablenotetext{*}{Assumes zero albedo and full heat redistribution.}
\tablenotetext{\dagger}{95\% credible upper limit.}
\end{deluxetable*}

\begin{deluxetable*}{lCC}[thb]
\tablecaption{Parameter distributions used for N-body simulations. \label{tab:distributions}}
\tablehead{
    \colhead{Parameter} &
    \colhead{Distribution} &
    \colhead{Distribution} \\
    \colhead{} &
    \colhead{\textit{(3-body)}} &
    \colhead{\textit{(4-body)}}
}
\startdata
\textit{\textbf{Planet f}} & &  \\
Mass, $M_\text{f}$ (\Mearth) & 12.3 & \mathcal{T}(12.3, 3.1, 0.0, \infty) \\
Orbital period, $P_\text{f}$ (days) & 542.0798 & \mathcal{N}(542.0798, 0.0001) \\
Eccentricity, $e_\text{f}$ & 0.035 & \mathcal{T}(0.005, 0.005, 0.0, 1.0) \\
Mean anomaly, $\mathcal{M}_\text{f}$ & \mathcal{U}(0, 2\pi) & \mathcal{U}(0, 2\pi) \\
Argument of periastron, $\omega_\text{f}$ ($^\circ$) & 0.0 & \mathcal{T}(231, 120, 0.0, 360) \\
Inclination, $i_\text{f}$ ($^\circ$) & 90.0 & \mathcal{T}(89.97, 0.01, 0.0, 90.0) \\
\hline
\textit{\textbf{Planet e}} & &  \\
Mass, $M_\text{e}$ (\Mearth) & \text{N/A} & \mathcal{T}(12, 5, 0.0, \infty) \\
Orbital period, $P_\text{e}$ (days) & \text{N/A} & [344, 394) \\
Eccentricity, $e_\text{e}$ & \text{N/A} & [0.0, 0.3) \\
Mean anomaly, $\mathcal{M}_\text{e}$ & \text{N/A} & \mathcal{U}(0, 2\pi) \\
Argument of periastron, $\omega_\text{e}$ ($^\circ$) & \text{N/A} & \mathcal{T}(114, 55, 0, 360) \\
Inclination, $i_\text{e}$ ($^\circ$) & \text{N/A} & \mathcal{T}(89.84, 0.05, 0.0, 90.0) \\
\hline
\textit{\textbf{Satellite}} & &  \\
Mass, $M_\text{s}$ (\Mearth) & 0.15 & 0.15 \\
Semi-major axis, $a_\text{s}$ ($R_\text{H}$) & [0.1, 0.8) & \mathcal{U}(0.1, 0.4) \\
Eccentricity, $e_\text{s}$ & 0.0 & 0.0 \\
Mean anomaly, $\mathcal{M}_\text{s}$ & \mathcal{U}(0, 2\pi) & \mathcal{U}(0, 2\pi) \\
Argument of periastron, $\omega_\text{s}$ ($^\circ$) & 0.0 & 0.0 \\
\tablenotemark{a}Cosine of inclination, $\cos (i_\text{s})$ & [0, 1) & \mathcal{U}(0, 1) \\
\enddata
\tablecomments{$\mathcal{N}(\mu, \sigma)$ is a normal distribution with mean $\mu$ and standard deviation $\sigma$; $\mathcal{U}(a, b)$ is a uniform distribution bounded between $a$ and $b$; $\mathcal{T}(\mu, \sigma, a, b)$ is a truncated normal distribution with mean $\mu$ and standard deviation $\sigma$ and bounded between $a$ and $b$.}
\tablenotetext{a}{The inclination of the satellite's orbit, $i_\text{s}$, is defined from the reference plane of the planet's orbit.}
\end{deluxetable*}

\section{Methods} \label{sec:methods}

We first consider whether it is feasible that \HIP~could host an exomoon over the system's lifetime. Here we outline a theoretical approach to investigating this possibility. First, we apply a suite of dynamical simulations to assess the long-term orbital stability of satellites around \HIP. We then separately consider the effects of tidal friction on the orbits of satellites and whether tidal migration can further destabilize the system. Throughout this section, we assume that \HIP~has properties consistent with measurements from \citet{Santerne+2019}, and we do not self-consistently include circumplanetary rings in our simulations.

\subsection{N-body Simulations}\label{sec:methods:nbody}

\subsubsection{Three-body Simulation}\label{sec:methods:nbody:threebody}

Because HIP\,41378 is a multiplanet system, gravitational interactions between neighboring planets may affect the long-term stability of a satellite orbiting planet f. Before considering the effects of additional planets, we start with an idealized three-body system with only one satellite orbiting planet f, which is in orbit around HIP\,41378. To simulate the dynamical evolution of the system, we use the \texttt{REBOUND} N-body code \citep{rebound} and the 15th order ``IAS15'' integrator with adaptive time stepping \citep{reboundias15}.

In our simulations, we fix the stellar mass and the mass and semi-major axis of planet f to the values from \citet{Santerne+2019}, provided in Table \ref{tab:planets}. Given a vast range of possible satellite configurations and properties, which would be computationally prohibitive to fully explore, we chose to consider a satellite with the largest satellite-to-planet mass ratio observed in the Solar System (i.e., the Moon-Earth system). This choice allows us to explore a physically motivated scenario that maximizes the potential signal produced by the satellite---larger satellites may be exceedingly rare, based on the population of hundreds of moons in the Solar System, and smaller moons are far less likely to be detected. Therefore, as an extreme case, we consider a satellite of \HIP~with a mass of $M_\text{s} = 0.15$~\Mearth~to be consistent with the satellite-to-planet mass ratio for the Earth-Moon system ($M_\text{M}/M_\oplus \approx 0.0123$). We prescribe the satellite an Earth-like rocky composition ($\rho_\text{s} = 5.5 \,\text{g cm}^{-3}$), yielding a radius of $R_\text{s} = 0.53$~\Rearth~(approximately the size of Mars).

For simplicity, the moon is initiated on a circular orbit around planet f, then, following the results of previous studies estimating stability limits of hierarchical systems \citep[e.g.,][]{Rosario-Franco+2020AJ, Quarles+2021AJ, Jagtap+2021}, each simulation is run for a duration of up to $10^5$ yr. A simulation is halted and marked as ``unstable'' if either of the following criteria are met:
\begin{itemize}
    \item the satellite collides with the planet\footnote{We also considered imposing an inner stability bound defined by the planet's Roche limit radius, $R_\text{Roche}~\approx~2.4R_\text{s}(M_\text{p}/M_\text{s})^{1/3}$. However, due to the planet's low observed mass, the conventional definition of the Roche limit yields a nonphysical distance smaller than the planet's measured radius. Without further information about the planet's interior structure or whether it possesses a circumplanetary ring, we chose to set the inner separation bound equal to the planet's observed radius instead.}, or 
    \item the satellite escapes the gravitational influence of the planet by crossing outside of the planet's Hill radius, defined as
        \begin{equation}\label{eq:hill}
            R_\text{H} = a_\text{p} \bigg( \frac{M_\text{p}}{3 M_\star} \bigg)^{1/3} ,
        \end{equation}
        where $a_\text{p}$ is the planet's semi-major axis, and $M_\text{p}$ and $M_\star$ are the masses of the planet and star, respectively.
\end{itemize}

As a test case, we first reproduced the results of \citet{Domingos+2006} and \citet{Rosario-Franco+2020AJ} by assessing the stability of the hypothetical satellite as a function of the initial planet-moon semi-major axis, $a_\text{s}$, and the eccentricity, $e_\text{f}$, of the host planet's orbit. Because the eccentricity of planet f is well-constrained by observations (see Table \ref{tab:planets}), the main purpose of this setup was to benchmark our N-body code and confirm that it generates expected results---we provide a more thorough description of this procedure in Appendix \ref{app:three-body}. In summary, we qualitatively recover a similar trend in the stability of the exomoon as a function of $(a_\text{s}, e_\text{f})$ as previous studies, and the stability limit we recover for circular planetary orbits, $a_\text{crit} \approx 0.41 R_\text{H}$, is consistent with published values \citep{Rosario-Franco+2020AJ}.

Next, we investigate how the satellite's orbital stability depends on its initial inclination. Our setup is similar to that of our benchmark simulations and \citet{Domingos+2006} and \citet{Rosario-Franco+2020AJ}, but here we fix planet f's eccentricity to its measured 95\% confidence upper limit of $e_\text{f} = 0.035$ \citep{Santerne+2019}, while varying the satellite's orbital inclination, $i_\text{s}$, and semi-major axis, $a_\text{s}$. We also tested a zero-eccentricity case and found similar results (in general, lower-eccentricity systems are expected be more stable regardless of satellite inclination; see Appendix \ref{app:three-body}).

We constrained our simulations to vary $i_\text{s}$ from $\cos i_\text{s}=0$ (aligned with the planet's reference plane) to $\cos i_\text{s}=1$ (perpendicular to the planet's reference plane), and $a_\text{s}$ from 0.1$R_\text{H}$ to 0.8$R_\text{H}$. Here, $R_\text{H}$ is the Hill radius defined in Equation \ref{eq:hill} ($R_\text{H} \approx 0.03$ AU $\approx 76 R_\text{p}$). The planet and satellite are both initialized with their ascending node longitudes and pericenter arguments set to zero ($\Omega = \omega = 0^\circ$), and the mean anomaly of the planet is also set to zero ($\mathcal{M}_\text{f} = 0^\circ$).

Then, to efficiently explore the moon's stability within our defined parameter space, we run a suite of N-body simulations using a quadtree data structure implemented in \texttt{astroQTpy}\footnote{\url{https://github.com/CalebHarada/astroQTpy}} \citep{harada_astroQTpy_2023}, which is based on the transit injection and recovery quadtree algorithm described by \citet{Ment+2023}. We initially create a $4 \times 4$ stability grid by evenly subdividing the parameter space in $\cos i_\text{s}$ and $a_\text{s}$. Within each grid cell, we run 25 independent N-body simulations\footnote{We experimented using higher numbers of simulations per grid cell, but found that this has no significant impact on our results.} with the initial conditions $\cos i_\text{s}$ and $a_\text{s}$ drawn from uniform distributions. The satellite's initial mean anomaly, $\mathcal{M}_\text{s}$, is also chosen randomly from a uniform distribution for each simulation (see Table \ref{tab:distributions}). Each simulation is run forward for up to $10^5$ yr, and then the fractional stability of each cell, $f_\text{stab}$, is calculated as the fraction of simulations (out of 25) which satisfy our stability criteria defined above.

In this setup, each grid cell corresponds to a child node of the quadtree which occupies $(1/4) \times (1/4)$ of the total initial parameter space $\{\cos i_\text{s}\} \times \{a_\text{s}\}$. The grid is then iteratively refined to higher resolution by splitting each node into four equal child nodes where the following criteria are satisfied:
\begin{itemize}
    \item the fractional stability of the node differs from that of a neighboring node by at least 20\%\footnote{We replicated our simulations using a more stringent threshold of 10\%, but found that this had no major impact on our results.}, and
    \item each of the resulting child nodes occupies at least $(1/32) \times (1/32)$ of the total parameter space.
\end{itemize}

\begin{figure*}[tb!]
    \centering
    \includegraphics[width=0.8\textwidth]{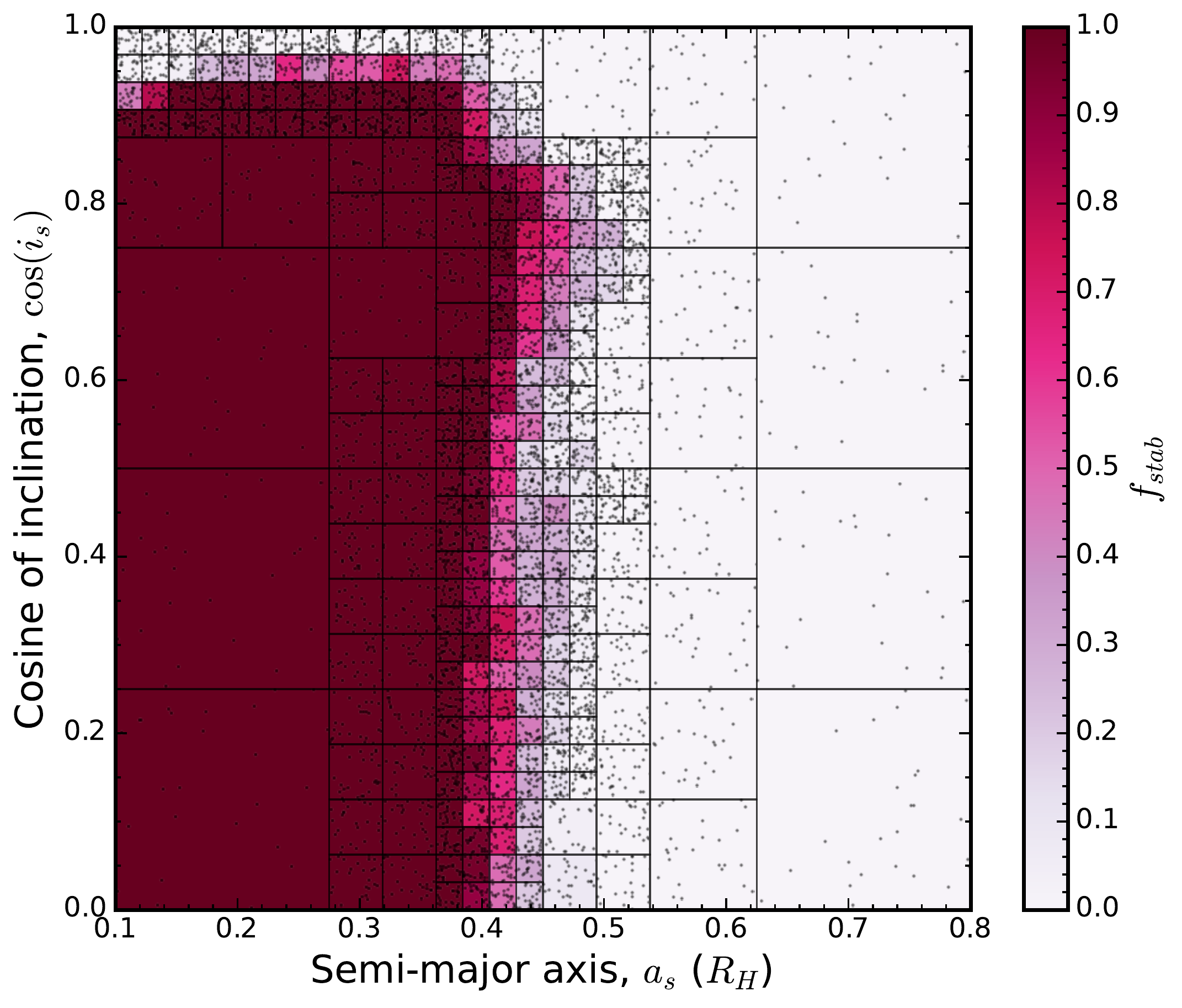}
    \caption{Three-body orbital stability map for a system consisting of the star HIP\,41378, planet f, and planet f's hypothetical satellite, as a function of the satellite's semi-major axis $a_\text{s}$ (expressed in terms of the Hill radius, $R_\text{H}$) and the cosine of the satellite's inclination $\cos i_\text{s}$ (with respect to the planet's reference plane). The color of each grid cell corresponds to the fraction $f_\text{stab}$ of the 25 simulations contained inside the cell (shown as points) which survive for $10^5$ yr. Absolutely unstable regions are represented by the lightest areas, while absolutely stable regions are shown by the darkest red regions. Note that moons with high orbital inclinations ($\cos i_\text{s} \gtrsim 0.8$) tend to become unstable due to Kozai-Lidov oscillations, and there are no stable orbits for $\cos i_\text{s} \gtrsim 0.95$.}
    \label{fig:nbody-a_moon-vs-inc_moon}
\end{figure*}

After a node has split, all of the completed simulations stored within that node are distributed to its four child nodes according to their $\cos i_\text{s}$ and $a_\text{s}$ values, and each child node is subsequently filled by running more N-body simulations until each node contains a maximum of 25 simulations. This process is repeated until the splitting criteria are no longer true for any nodes. Therefore, the total number of N-body simulations run for the completed stability map is between $4 \times 4 \times 25 = 400$ and $32 \times 32 \times 25 = 25,600$. Figure \ref{fig:nbody-a_moon-vs-inc_moon} shows the final stability map for this setup, which we will discuss further in Section \ref{sec:results_disc}.

\subsubsection{Four-body Simulation}

We next run an additional set of N-body simulations, now including the second-outermost planet in the system, HIP\,41378\,\,e (which is separated from planet~f by $\sim$10 mutual Hill radii). It is beyond the scope of this work to simulate the global stability of the system, so we do not include planet b ($P=15.57$ d), planet c ($P=31.71$ d), or planet d ($P=278.36$ d) in our simulations in order to reduce the complexity of the problem and lower computational costs. For the purpose of studying the stability of a satellite around planet f, this is a valid assumption given that the force of gravity on planet f due to the three inner planets at closest approach is orders of magnitude smaller than that of planet e (planets b, c, and d are separated from planet f by approximately 65, 61, and 19 mutual Hill radii, respectively). Moreover, the transit timing variations of planet f are thought to be dominated by a 2:3 period commensurability with planet e \citep{Bryant+2021}.

For this case, our setup is similar to that of our suite of three-body simulations, but we now evaluate the fractional stability as a function of planet e's orbital period, $P_\text{e}$, and eccentricity, $e_\text{e}$. Furthermore, to account for observational uncertainties in the masses and orbits of planets e and f, we randomly draw other system parameters that are consistent with the results of \citet{Santerne+2019}. For each trial simulation we randomly draw the mass, argument of pericenter, and inclination of planets e and f, and the period and eccentricity of planet f from normal distributions with standard deviations consistent with the uncertainties given in Table \ref{tab:planets}. 

\begin{figure*}[thb!]
    \centering
    \includegraphics[width=0.95\textwidth]{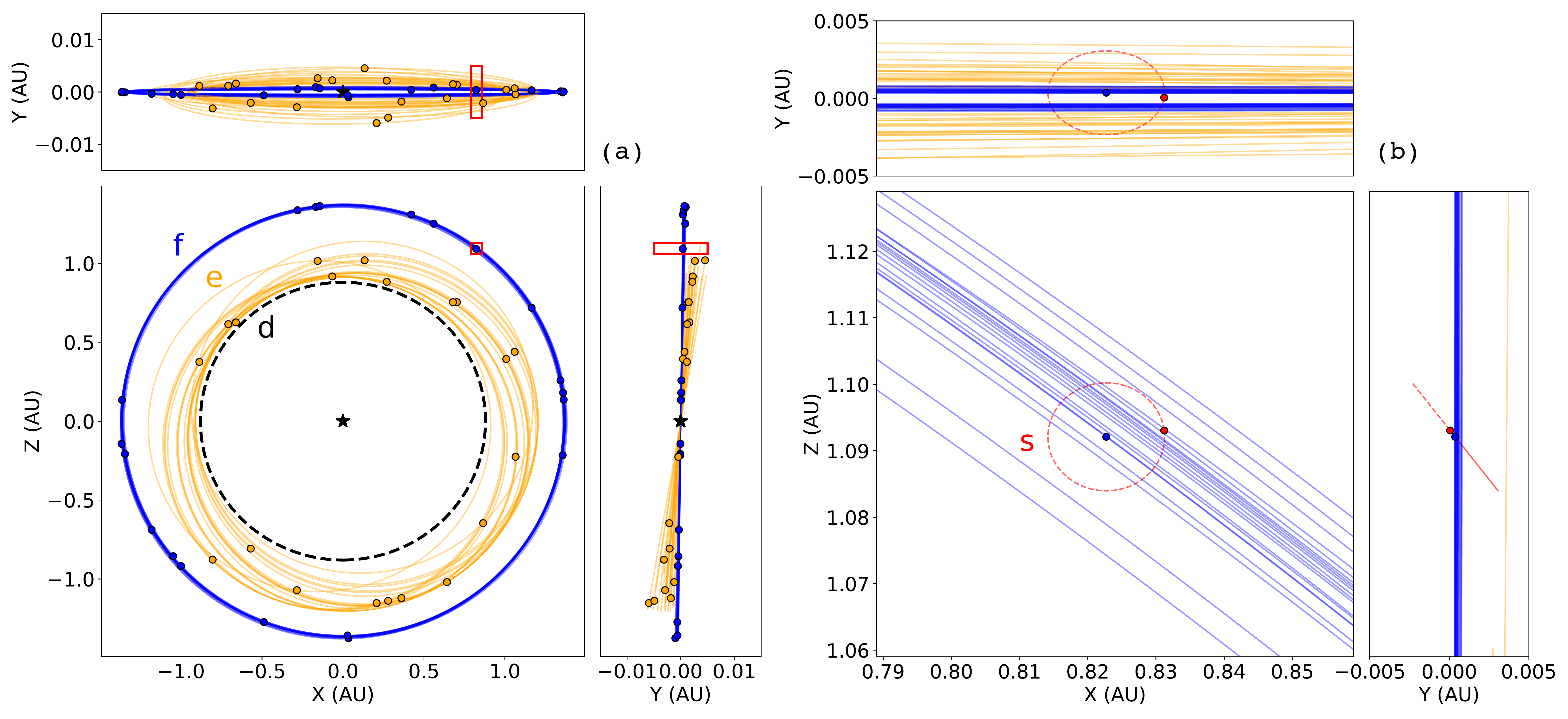}
    \caption{Example two-dimensional projections of the initial positions and orbits of each object in our four-body simulations, including planet e. Panel (a) shows the top-down and edge-on projections of 20 random realizations of the system, with the initial orbits (lines) and positions (points) of planet e shown in orange, and the initial orbits and positions of planet f shown in blue. The black ``$\star$'' symbol shows the position of the host star, and the dashed black line shows the (assumed) fixed orbital semi-major axis for planet d. The small red boxes indicate the limits of the area plotted in panel (b), which shows a zoomed-in version of the system highlighting the initial position and orbit of the satellite, marked ``s'' (red), orbiting planet f.}
    \label{fig:nbody-setup}
\end{figure*}

For the satellite of planet f, we assume a circular orbit and randomly draw the cosine of the inclination (with respect to the planet's orbital plane) from a uniform distribution between $0$ and $1$, and the initial semi-major axis between 0.1 and 0.4$R_\text{H}$ (i.e., within the stability limits for a low-eccentricity host planet derived from our benchmark simulations). We assume the same fixed stellar mass and satellite mass as before, and initialize each object with a random mean anomaly between $0^\circ$ and $360^\circ$. A summary of the distributions used to select parameters for each iteration is provided in Table \ref{tab:distributions}. Figure \ref{fig:nbody-setup} shows an example of 20 random realizations of the system setup for fixed values of $P_\text{e}$ and $e_\text{e}$. 

With the addition of planet e to our simulations, we update the stability criteria defined in Section \ref{sec:methods:nbody:threebody}, such that a system is also flagged as ``unstable'' if planet e enters the Hill sphere of planet f, or if any planet's orbit crosses the orbit of another planet. The latter criterion places an upper limit on the eccentricity of planet e for a given orbital period due to the intersection of its periastron distance with the expected orbit of planet d (note, however, that we do not explicitly include planet d in the N-body simulations). Assuming that planet d has a circular orbit\footnote{This is consistent with \citet{Santerne+2019}, who report a measured eccentricity for planet d of $e_\text{d} = 0.06 \pm 0.06$.}, this limit is set by
\begin{equation}\label{eq:crit_p}
    P_\text{e,crit}(e_\text{e}) \simeq \frac{2 \pi}{\sqrt{G M_\star}} \bigg(\frac{a_\text{d}}{1 - e_\text{e}}\bigg)^{3/2}
\end{equation}
where $a_\text{d}$ is the semi-major axis of planet d and $G$ is the Newtonian gravitational constant. Note that for the measured eccentricity of planet e \citep[$e_\text{e} = 0.14$;][]{Santerne+2019}, $P_\text{e,crit} \approx 351 \,\text{d}$, which is only within $\sim$1.8$\sigma$ of the measured period of planet e \citep[$P_\text{e} = 369 \pm 10 \,\text{d}$;][]{Santerne+2019}.

\begin{figure*}[tb!]
    \centering
    \includegraphics[width=0.8\textwidth]{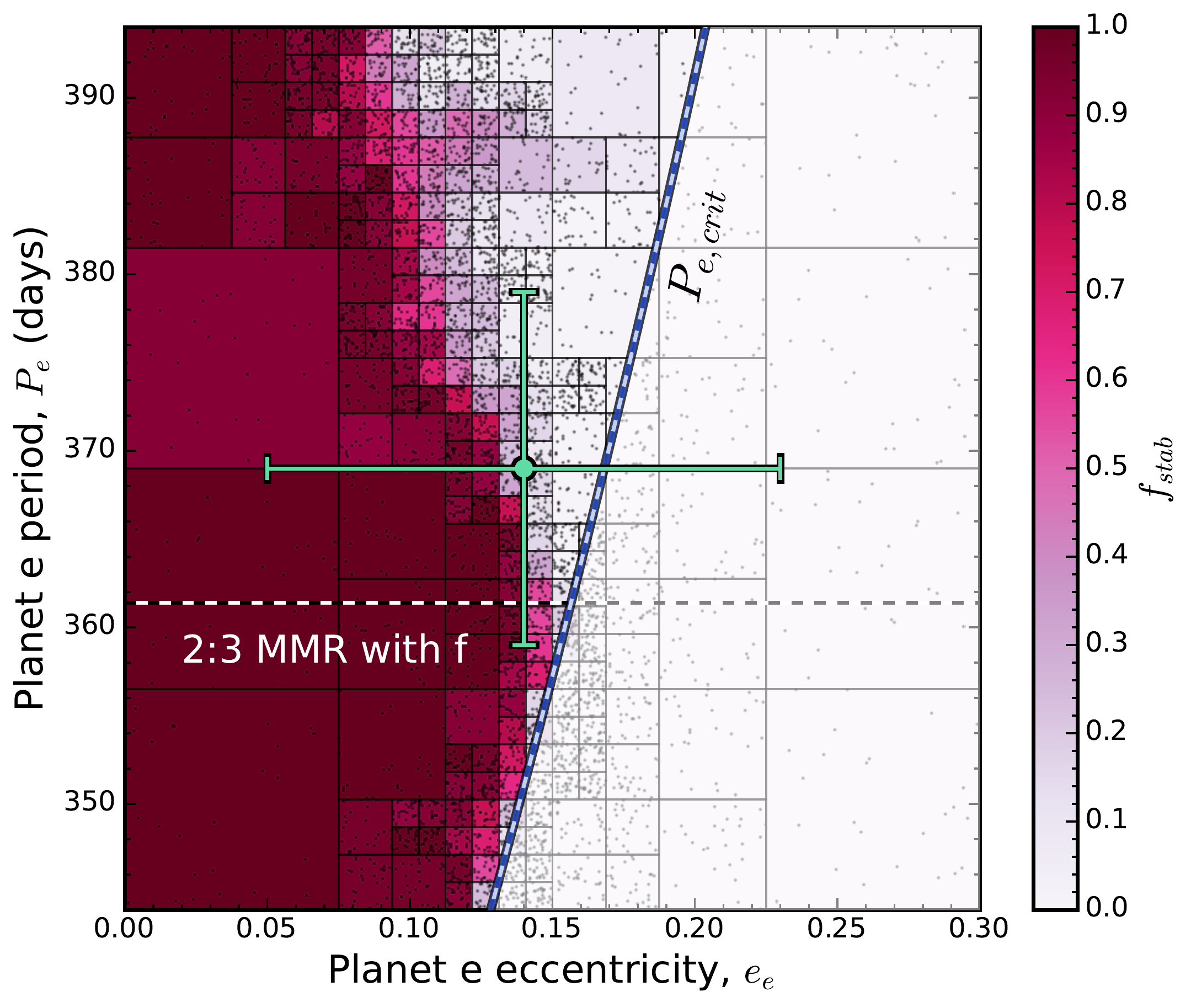}
    \caption{Four-body orbital stability map for a system including the star HIP\,41378, planets e and f, and a satellite orbiting planet f, as a function of planet e's initial orbital period ($P_\text{e}$) and eccentricity ($e_\text{e}$). As in Figure \ref{fig:nbody-a_moon-vs-inc_moon}, the color of each grid cell shows the fraction of 25 simulations which are stable over $10^5$ yr. Here, the dashed blue line shows $P_\text{e,crit}(e_\text{e})$, the theoretical upper limit on planet e's eccentricity vs.~period due to the orbit of planet d (see Equation \ref{eq:crit_p})---the region to the right of this line is unstable by definition because planet e would cross the orbit of planet d. The measured period and eccentricity of planet e from \citet{Santerne+2019} are shown by the green point with errorbars representing 1-$\sigma$ uncertainties. For reference, the period of planet e corresponding to a 2:3 mean motion resonance (MMR) with planet f is shown by the horizontal dashed line.}
    \label{fig:nbody-per_e-vs-ecc_e}
\end{figure*}

We explore the fractional stability of the system in the parameter space around this limit over a timescale of $10^5$ yrs, selecting a range of $P_\text{e}$ and $e_\text{e}$ consistent with their observed values within $\sim$2$\sigma$ \citep[see Table \ref{tab:distributions};][]{Santerne+2019}. Again, we start by evenly dividing the entire $(P_\text{e}, e_\text{e})$ parameter space into a $4 \times 4$ grid, and computing $f_\text{stab}$ in each cell by running a set of 25 independent trials with the set of parameters described above. We then proceed by iteratively subdividing the quadtree according to the criteria defined in Section \ref{sec:methods:nbody:threebody} and computing more N-body simulations until no more cells can be split. The resulting stability map is shown in Figure \ref{fig:nbody-per_e-vs-ecc_e}.

\subsection{Tidal Migration}

In addition to gravitational interactions with other planets, tidal interactions can lead to the secular evolution of a moon's orbit. Here we consider the possible effects of tides on the orbital stability of moons around \HIP. Here, we implement an equilibrium tide (ET) model, which assumes a simple parameterization to capture the relevant tidal processes without considering the detailed physics of the interiors of the bodies or 3D effects \citep[see][and references therein]{Barnes_2017}. In summary, the ET model assumes that the gravitational force of a satellite creates an elongated bulge in the planet in a state of equilibrium, whose long axis is misaligned with the line connecting the planet and moon centers of mass due to dissipative processes within the planet. This misalignment, or ``lag'' between the passage of the tidal bulge and the satellite, creates torques which lead to the secular evolution of the spin and orbital angular momenta of the two bodies. 

As detailed in \citet{Barnes_2017}, two distinct parameterizations of the tidal lag are well established in the literature: the so-called ``constant time lag'' (CTL) model, in which the time interval between the passage of the moon and the tidal bulge is fixed, and the ``constant phase lag'' (CPL) model in which the phase between the moon and the tidal bulge is constant, independent of frequency. In both frameworks, the energy dissipation and tidal bulge are coupled via two parameters: the tidal Love number of degree 2, $k_2$, which accounts for the internal forces of the deformed body (where $k_2=0$ describes a perfectly rigid body and $k_2=3/2$ a perfectly fluid body), and a parameter representing the offset between the tidal bulge axis and the line connecting the centers of mass. In the CTL model the latter parameter is referred to as the ``tidal time lag,'' $\tau$, which is inversely related to the tidal quality factor $Q$ in the CPT model. These parameters effectively describe the efficiency of the tidal response between the two bodies---larger values of $\tau$ (smaller $Q$) imply faster tidal evolution. While both CTL and CPL generally produce qualitatively similar results for the Solar System, one may be more suitable than the other depending on the specific scenario. In this work we choose to implement the CTL model, as it allows us to explore the impact of planetary rotation rate on the tidal evolution of the moon.

In both ET model frameworks, the tidal evolution of two bodies is described by a set of six differential equations and six free parameters: the semi-major axis $a$, the eccentricity $e$, the rotation rates of the primary and secondary bodies $\Omega_i$, and their two obliquities $\psi_i$ (where $i=1,2$ correspond to the primary and secondary bodies). Following \citet{Barnes_2017} and references therein \citep{Mignard_1979, Hut_1981, Greenberg_2009, Leconte+2010A&A, Heller+2011A&A}, the evolution for the CTL model is described by the following set of differential equations:
\begin{multline}
    \frac{de}{dt} = \frac{11 a e}{2 G M_1 M_2} \\ \times \sum_{i=1}^2 Z^\text{ctl}_i \bigg( \cos(\psi_i) \frac{f_4(e)}{\beta^{10}(e)}\frac{\Omega_i}{n} - \frac{18}{11} \frac{f_3(e)}{\beta^{13}(e)} \bigg),
\end{multline}
\begin{multline}\label{eq:ctl_da_dt}
    \frac{da}{dt} = \frac{2a^2}{G M_1 M_2} \\ \times \sum_{i=1}^2 Z^\text{ctl}_i \bigg( \cos(\psi_i) \frac{f_2(e)}{\beta^{12}(e)}\frac{\Omega_i}{n} - \frac{f_1(e)}{\beta^{15}(e)} \bigg),
\end{multline}
\begin{multline}
    \frac{d\Omega_i}{dt} = \frac{Z^\text{ctl}_i}{2 M_i r^2_{g,i} R^2_i n } \\ \times \bigg( 2 \cos(\psi_i) \frac{f_2(e)}{\beta^{12}(e)} - \big[ 1 + \cos^2(\psi_i) \big] \frac{f_5(e)}{\beta^9(e)} \frac{\Omega_i}{n} \bigg),
\end{multline}
and
\begin{multline}
    \frac{d\psi_i}{dt} = \frac{Z^\text{ctl}_i \sin(\psi_i)}{2 M_i r^2_{g,i} R^2_i n \Omega_i } \\ \times \bigg( \bigg[ \cos(\psi_i) - \frac{\xi_i}{\beta} \bigg] \frac{f_5(e)}{\beta^{9}(e)} \frac{\Omega_i}{n} - 2 \frac{f_2(e)}{\beta^{12}(e)} \bigg)
\end{multline}
where $t$ is time, $M_i$ are the masses of the two bodies, $R_i$ are their radii, and $n$ is the mean motion. The quantity $r_g$ is the radius of gyration, which is related to the moment of inertia by $I = M (r_g R)^2$. The above equations are averaged over the orbital period and therefore represent the mean variation of the orbital parameters. 

The factors $Z^\text{ctl}_i$ and $\xi_i$ are given by
\begin{gather}
    Z^\text{ctl}_i \equiv 3 G^2 k_{2,i} M^2_j (M_i + M_j) \frac{R^5_i}{a^9} \tau_i, \\
    \xi_i \equiv \frac{r^2_{g,i} R^2_i \Omega_i a n}{G M_j}
\end{gather}
where $k_{2,i}$ are the second order tidal Love numbers, $\tau_i$ are the tidal time lags, and the subscripts $i$ and $j$ refer to the two bodies. The remaining quantities are given by:
\begin{align*}
    \beta(e) &= \sqrt{1 - e^2}, \\
    f_1(e) &= 1 + \frac{31}{2}e^2 + \frac{255}{8}e^4 + \frac{185}{16}e^6 + \frac{25}{64}e^8 , \\
    f_2(e) &= 1 + \frac{15}{2}e^2 + \frac{45}{8}e^4 + \frac{5}{16}e^6 , \\
    f_3(e) &= 1 + \frac{15}{4}e^2 + \frac{15}{8}e^4 + \frac{5}{64}e^6 , \\
    f_4(e) &= 1 + \frac{3}{2}e^2 + \frac{1}{8}e^4 , \\
    f_5(e) &= 1 + 3e^2 + \frac{3}{8}e^4 .
\end{align*}

Unlike hot Jupiters ($P \lesssim 10$ days), which are expected to be in tidally synchronous rotation states with their host stars, we expect the orbital separation of \HIP~to be large enough such that tidal effects from its host star are negligible. We test this assumption using a CTL equilibrium tide model as implemented in the open source software package \texttt{EqTide} \citep{Barnes_2017}. 

We assign the host star a mass, radius, and rotation period consistent with \citet{Santerne+2019}, and a solar radius of gyration, $r_{g,\star} \approx 0.26$ \citep[e.g.,][]{Claret+1989AAS}. We assume a Love number of $k_{2, \star} = 0.5$ and constant time lag of $\tau_\star = 0.01$ s, following \citet{Barnes_2017}. For \HIP, we use the planet mass and radius reported in \citet{Santerne+2019} and assume a rotation period of 10 hours\footnote{Note that faster rotation will generally lead to tidal migration on shorter timescales, which we explore later in this section.}, similar to that of the Solar System gas giants. For the radius of gyration and Love number \citep[which are generally not measurable for exoplanets except in special cases, e.g.,][]{Hellard+2019ApJ, Akinsanmi+2019A&A, Barros+2022A&A}, we assume analogous values to Jupiter's, and use recent measurements of Jupiter from the Juno spacecraft, $r_{g, \text{f}}=0.52$ \citep{Ni_2018} and $k_{2, \text{f}} = 0.565$ \citep{Idini+2021}. Note that Saturn has similar properties, but a slightly smaller Love number of $k_2 = 0.390$ \citep{Lainey+2017}, which would not significantly affect our results. For the constant time lag, we also assume an analogous value to Solar System gas giants, $\tau_\text{f} = 0.00766$ s \citep[e.g.,][]{Tokadjian+2020}. For reference, note that $\tau \approx 0.766$ s for Neptune-like planets and $\tau \approx 638$ s for rocky planets \citep{Tokadjian+2020}. 

Under these assumptions, we evolve the CTL model forward for a duration of $3.1 \times 10^{9}$ yr \citep[the approximate age of the system;][]{Santerne+2019}, starting the planet with its currently measured semi-major axis and orbital eccentricity. We find that the relative change in semi-major axis for planet f is $\ll 1$ ppm, and the change in rotation period is less than 0.1\%. Therefore, tidal interactions between planet f and its host star are negligible and we can safely ignore the tidal influence of the star for the purposes of this study.

Next, ignoring the influence of tides from the host star and any other planets, we investigate the tidal stability of a hypothetical moon orbiting planet f. As in Section \ref{sec:methods:nbody}, we consider the extreme case of an Earth-composition satellite with the same mass ratio to planet f as the Moon-Earth system ($M_\text{s} = 0.15$ \Mearth, $R_\text{s} = 0.53$ \Rearth). Again, this choice is motivated by the largest moon mass ratio in the Solar System and the fact that smaller moons would be much harder to measure. For the tidal properties, we arbitrarily assume that the satellite has a gyration radius similar to the measured value for the Solar System's largest moon $r_{g,\text{s}} = 0.56$ \citep[Ganymede;][]{Showman+1999}, and a tidal Love number and constant time lag consistent with measurement of rocky planets in the Solar System, $k_{2,\text{s}} = 0.3$ and $\tau_\text{s} = 638$ s \citep[e.g.,][and references therein]{Tokadjian+2020}. The mass and radius of the planet are again assumed to be the values reported in \citet{Santerne+2019}, and the gyration radius and Love number are set to the respective Jovian values \citep{Ni_2018, Idini+2021}. 

\begin{figure*}[th!]
    \centering
    \includegraphics[width=0.95\textwidth]{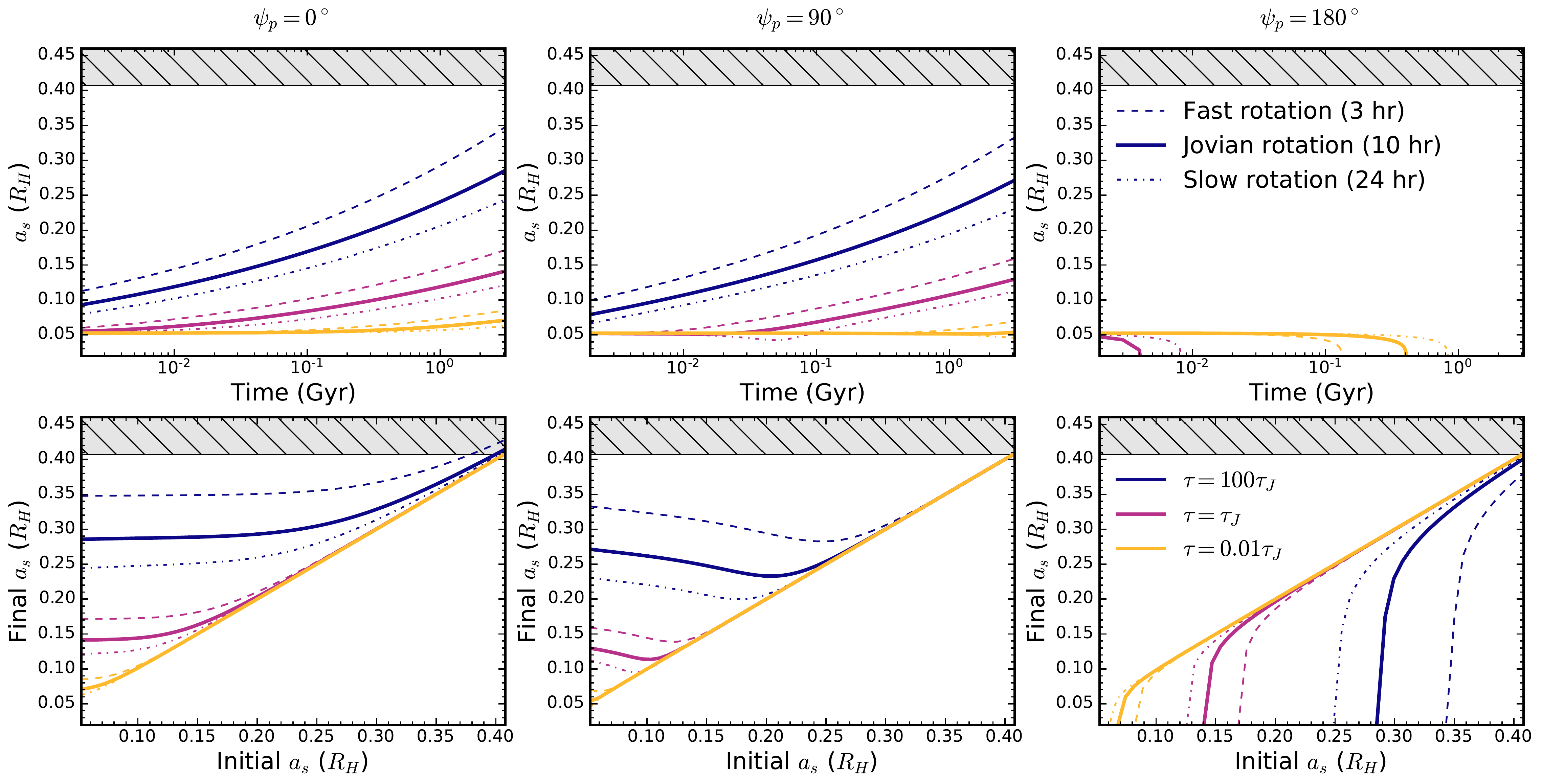}
    \caption{Equilibrium tide evolution models for a 0.15 \Mearth (0.53 \Rearth) satellite orbiting \HIP~computed with the CTL mode in \texttt{EqTide} \citep{Barnes_2017} for different planet obliquities. \textbf{Top row:} Moon semi-major axis as a function of time over the lifetime of the system, starting from an initial orbital separation of 4$R_p$. \textbf{Bottom row:} Final vs.~initial moon semi-major axis after a timescale of $3.1$ Gyr, the current estimated age of the system \citep{Santerne+2019}. The columns from left to right show the results for different assumed planet obliquities (moon inclinations): $\psi_\text{p} = 0^\circ$ (aligned prograde orbit), $\psi_\text{p} = 90^\circ$ (polar orbit), and $\psi_\text{p} = 180^\circ$ (aligned retrograde orbit). Each model assumes a different constant time lag $\tau$ for the planet, indicated by color (where $\tau_\text{J} = 0.00766$ s is roughly Jupiter's value). The fiducial planet rotation period is assumed to be 10 hours (solid lines), but we also consider a slow rotation case ($P_\text{rot} = 24$ hours; dashed lines) and a fast rotation case ($P_\text{rot} = 3$ hours; dot-dashed lines). The gray hatched regions highlight regions where the moon tends to become dynamically unstable, $a_\text{s} \gtrsim 0.41\,R_\text{H}$) (see Figure \ref{fig:nbody-a_moon-vs-inc_moon}). Note that the ratio of the final-to-initial moon semi-major axis approaches unity as the starting semi-major axis increases in each obliquity case we considered. This is consistent with the very strong dependence of $\tau_\text{mig}$ on the semi-major axis in Equation \ref{eq:timescale} for the zero-obliquity case (i.e., $\tau_\text{mig} \propto a^8$). Note also that for the retrograde ($\psi_\text{p} = 180^\circ$) case, moons with initial orbits close to the planet tend to migrate inward and ultimately collide with the planet, depending on the assumed planetary rotation and tidal lag constant.}
    \label{fig:tidal}
\end{figure*}

Because the obliquity, constant time lag, and rotation period of the \HIP~are almost entirely unconstrained, especially given the low bulk density inferred from mass and radius measurements, we explore how changing these parameters affects the final semi-major axis of the moon after a duration of $3.1$ Gyr. Again using the \texttt{EqTide} code \citep{Barnes_2017}, we evolve the system forward in time starting the satellite on a tidally-synchronous, circular orbit. We calculate a suite of models, varying the initial semi-major axis of the moon from 4$R_p$ ($\approx$0.0525$R_\text{H}$) to 0.4$R_\text{H}$ (i.e., the approximate stability limit for a circular orbit of planet f from our N-body simulations). As an initial case, we choose $\tau_\text{f}$ to be consistent with the estimates for the Solar System gas giants, $\tau_\text{J} = 0.00766$ s \citep{Tokadjian+2020} and a rotation period similar to that of Jupiter, $P_\text{rot,f} = 10$ hr. Then, we investigate the effect of planetary rotation rate by running the same grid of models assuming a slow planet rotation ($P_\text{rot,f} = 24$ hr) and a fast planet rotation ($P_\text{rot,f} = 3$ hr). For each case, we also test two extreme cases of the constant time lag by setting $\tau_\text{f}$ to 0.01 and 100 times the value assumed for Jupiter $\tau_\text{J}$. We then repeated each of these simulations assuming three different planet obliquities, $\psi_\text{p}$: $0^\circ$, $90^\circ$, and $180^\circ$. Lastly, after running each model for $3.1$ Gyr, we compare the final semi-major axis of the moon to the stability limit from dynamical simulations to determine whether tidal migration can significantly destabilize the system. Our results are shown in Figure \ref{fig:tidal}.

\section{Results and Discussion} \label{sec:results_disc}

In this section we discuss the results of our numerical N-body and tidal migration simulations. We then discuss the observability of large exomoons orbiting \HIP~in transit photometry by comparing theoretical finely-sampled light curves to observations from HST and K2, and then simulating idealized observations from the PLATO mission. Finally, we explore the potential impact of large moons with atmospheres on the observed transmission spectrum of \HIP.

\subsection{Stability Limits from Dynamical Simulations}

Our three-body simulations expanded upon the work of \citet{Rosario-Franco+2020AJ}, but in greater detail for the HIP\,41378 system. While \citet{Rosario-Franco+2020AJ} studied the stability of a Neptune-size moon orbiting a Kepler-1625b analog as a function of planet eccentricity and moon semi-major axis, here we simulated \HIP~with a 0.15 \Mearth~Mars-size satellite, varying the semi-major axis, $a_\text{s}$, inclination, $i_\text{s}$ and initial mean anomaly, $\mathcal{M}_\text{s}$, of the satellite. Ignoring the other planets in the system and assuming a fixed eccentricity of 0.035 \citep[95\% confidence upper limit;][]{Santerne+2019} for planet f, we defined the quantity $f_\text{stab}$ as the fraction of 25 simulations within each quadtree grid cell that are stable over $10^5$ yr. Figure \ref{fig:nbody-a_moon-vs-inc_moon} shows the results of our simulations, where $f_\text{stab}$ is quantified by color.

From the benchmark three-body simulations described in Appendix \ref{app:three-body}, the outer stability limit for a satellite on a co-planar orbit around a planet with an eccentricity of 0.035 is $a_\text{crit}~\approx~0.39~R_\text{H}$. Our results shown in Figure \ref{fig:nbody-a_moon-vs-inc_moon} show that this limit depends only weakly on the inclination of the satellite for $\cos i_\text{s} \lesssim 0.8$. At higher inclinations however, Kozai-Lidov oscillations can play a significant role in the dynamical evolution of the system, forcing the destabilization of the satellite over time. For these high-inclination systems, a trade off can occur between the satellite's inclination and eccentricity in order to conserve angular momentum, such that initially circular orbits become highly eccentric over time (i.e., bringing the satellite's pericenter distance closer to the planet, and the apocenter distance farther away). As shown in Figure \ref{fig:nbody-a_moon-vs-inc_moon}, this process starts to become important for $\cos i_\text{s} \gtrsim 0.8$, and dominates for $\cos i_\text{s} \gtrsim 0.95$ where there are no stable orbits.

The likelihood of these different inclination scenarios occurring in nature depends strongly on the moon's formation pathway. For moons formed within a circumplanetary disk, as is thought to be the case for the Galilean moon system around Jupiter, the inclination of the moon system relative to the planet spin axis should be low because of momentum conservation from the rotating circumplanetary disk. However, other formation scenarios, such as a kinetic impacts or dynamical capture, may result in moons with more misaligned inclinations, which we've demonstrated are less stable over long timescales, especially for closer-in orbits. We note, however, that tidal interactions may lead to significant outward migration of such close-in moons, thereby increasing their overall stability. We will discuss the effects of tidal migration in greater detail in Section \ref{sec:results_disc:tides}.

We then used the results from this three-body case and from our benchmark test in Appendix \ref{app:three-body} to inform our second suite of simulations, which more realistically represented the HIP\,41378 system with the inclusion of planet e. Since the observed orbit of planet f is nearly circular, for each simulation we randomly selected the initial semi-major axis of its satellite from a uniform distribution between $0.1 R_\text{H}$ and $0.4 R_\text{H}$, the approximate stability limit for low eccentricity planet orbits. The mass of the satellite was assumed to be the same as in our initial simulations (0.15 \Mearth), and the initial mean anomaly and inclination were drawn from uniform distributions. The other initial parameters for the satellite and planets e and f were chosen randomly from the distributions given in Table \ref{tab:distributions}. As before, we determined $f_\text{stab}$ by taking the fraction of 25 simulations within a given quadtree cell spanning $\{e_\text{e}\}$ and $\{P_\text{e}\}$ that survived over $10^5$ yr. Our results are shown in Figure \ref{fig:nbody-per_e-vs-ecc_e}, where again $f_\text{stab}$ is indicated by color. We also show the upper limit on $e_\text{e}$ as a function of $P_\text{e}$ (given by Equation \ref{eq:crit_p}), which is constrained by the (assumed fixed) orbit of planet d; and the observed values of $e_\text{e}$ and $P_\text{e}$ from \citet{Santerne+2019}.

The quasi-stable regions (left of the blue dashed line in Figure \ref{fig:nbody-per_e-vs-ecc_e}) are dominated by at least two independent effects. First, a trend of decreasing quasi-stability toward larger $e_\text{e}$ is set by gravitational interactions between planet e and planet f's satellite. At higher eccentricities, the apastron distance of planet e becomes larger, reducing the separation between its orbit and the orbit of planet f. Therefore, when planet e has a high eccentricity, there is a greater likelihood that it will gravitationally perturb planet f's satellite, leading to an unstable system. By the same physical reasoning, we also see a trend of decreasing quasi-stability along the period axis---for fixed eccentricity, the system becomes less stable as the period of planet e increases. In this case, the apastron distance also increases with period, according to Kepler's third law, which again subsequently increases the likelihood of planet e gravitationally perturbing planet f's satellite.

Second, the quasi-stable regions in Figure \ref{fig:nbody-per_e-vs-ecc_e} are also influenced by the randomly distributed parameters in each simulation, which are independent of the orbit of planet e. These manifest as random noise in the stability map. For example, in cases where the mass of planet f is $\gtrsim 2 \sigma$ less than the ``best-fit'' measured mass, the planet's Hill radius becomes sufficiently small such that moons with larger initial orbital separations can more easily be stripped away. Moreover, moons that start out with orbits highly inclined to the reference plane of the planet (especially those with smaller semi-major axes) are more likely to be unstable due to Kozai-Lidov oscillations.

Because the observed period and eccentricity of planet e reside in a transitional quasi-stable region of the map, we cannot robustly conclude whether a moon orbiting planet f could be stable given the uncertainties in planet e's measured period and eccentricity. Nonetheless, improved measurements of planet e's orbit may yet reveal that the system actually resides in a stable region of Figure \ref{fig:nbody-per_e-vs-ecc_e}, for instance, if its eccentricity is determined to be smaller than currently thought. If this turns out to be the case, then we would expect that most moons within $\sim 0.4 R_\text{H}$ of planet f should remain stable as long as their orbits are not highly inclined relative to the planet's reference plane. This result highlights the importance of modeling gravitational effects from other planets in multi-planet systems in the context of exomoon stability (and by extension, habitability), as well as the need for improved radial velocity measurements to measure planet masses in multi-planet systems.

\subsection{Limits from Tidal Evolution}\label{sec:results_disc:tides}

Tidal forces between two rotating bodies can lead to secular changes in their rotation and orbits. For example, a well-studied case in our Solar System is the gradual recession of Earth's Moon (and subsequent lengthening of the Earth's day) over Gyr timescales. We investigated how such tidal migration may influence the long-term orbital stability of a hypothetical moon orbiting \HIP. 

In our ET simulations outlined in Section \ref{sec:methods}, the secular evolution was parameterized by two intrinsic properties of the orbiting bodies, $k_2$ and $\tau$. Though both $k_2$ and $\tau$ are completely unknown for \HIP~(as are the moment of inertia and spin rate), there are constraints on these parameters for many Solar System bodies (though $\tau$ and $Q$ have uncertainties spanning orders of magnitude). We therefore assumed that the properties of \HIP~and any hypothetical moon are analogous to Solar System bodies. As the fiducial case, we assumed Jupiter-like values for $k_2$, $\tau$, $r_g$, and $\Omega$ for the planet, as described earlier in Section \ref{sec:methods}. We subsequently tested cases where $\tau$ was two orders of magnitude larger and smaller than the estimated Jovian value \citep[with the larger value representing a planet with Neptune-like properties; e.g.,][]{Tokadjian+2020}, and the rotation period was changed to either 3 hr (``fast'') or 24 hr (``slow''). Finally, we tested the effect of planet obliquity by repeating our simulations for $\psi_\text{p} = 0^\circ$, $\psi_\text{p} = 90^\circ$, and $\psi_\text{p} = 180^\circ$. For the satellite, we assumed tidally synchronous rotation, and $k_2$, $\tau$, and $r_g$ consistent with rocky Solar System bodies.

For simplicity, we assumed that each simulated moon started out with a circular orbit. We ran each simulation for 3.1 Gyr, varying the initial semi-major axis of the moon from 4$R_p$ ($\approx 0.0525 R_\text{H}$) to 30.5$R_p$ ($\approx 0.4 R_\text{H}$) such that the moon's initial orbital period was always longer than the planet's rotation period---this ensured that no inward migration would occur (for prograde satellites), in which case the moon could ultimately collide with the planet (e.g., for the 24 hr rotation case, the moon would migrate inward instead of outward if its semi-major axis were $\lesssim$1.6$R_p$). For each rotation rate, constant time lag, and obliquity case, we show an example of the time evolution of the satellite's semi-major axis in Figure \ref{fig:tidal}, as well as how the final semi-major axis of the satellite varies as a function of its initial semi-major axis.

Our results demonstrate that tidal migration does not significantly alter the stability of a moon orbiting \HIP~over the system's lifetime, except if the moon is in a retrograde orbit relative to the planet's spin ($\psi_\text{p} = 180^\circ$) or if $\tau$ or $\Omega$ of the planet is very large. In the former case, the moon can be forced to migrate inward sufficiently to collide with the planet, depending on the assumed tidal parameters and the planet's rotation rate (outward migration is not allowed here, so these results are independent of the N-body stability limit). Note that for the $\psi_\text{p} = 90^\circ$ case, some inward migration may occur depending on the initial conditions, but not sufficiently to cause a collision---however, as we showed with our N-body simulations, Kozai-Lidov oscillations play a significant role in the dynamical stability of high-inclination orbits. In fact, we expect that Kozai-Lidov oscillations would not permit a stable polar orbit ($\psi_\text{p} = 90^\circ$) scenario to begin with. We note also that, while we attempted to adequately represent the full dynamics of the system here, in nature we would expect both secular tidal migration and Kozai-Lidov oscillations to affect the stability of exomoons in a complex and coupled way. It is beyond the scope of our work here to self-consistently model both tides and orbital dynamics simultaneously.

For the prograde orbit case ($\psi_\text{p} = 0^\circ$), because the secular evolution of the moon's semi-major axis determines whether the moon can remain dynamically stable, it is useful for our discussion to introduce a timescale associated with the evolution of $a$, which we can write as $\tau_\text{mig} \sim a / \dot{a}$ (where $\dot{a}$ is the time derivative of $a$). Under the assumptions that $\psi_i = e = 0$, $M_\text{f} \gg M_\text{s}$, and $\Omega_\text{s}/n = 1$ (i.e., the moon is tidally locked), we can use the time derivative from Equation \ref{eq:ctl_da_dt} to show that the migration timescale scales as follows:
\begin{equation}\label{eq:timescale}
    \tau_\text{mig} \propto \frac{a^8}{k_{2,\text{p}} \tau_\text{p} M_\text{s} R_\text{p}^5} \bigg( \frac{\Omega_\text{p}}{n} - 1 \bigg)^{-1}
\end{equation}
where the subscript ``p'' indicates the planet (primary) and subscript ``s'' indicates the satellite (secondary). Note that in Equation \ref{eq:ctl_da_dt}, the assumption of setting $e=0$ implies that the $\beta(e)$, $f_1(e)$, and $f_2(e)$ terms become unity.

This scaling relation is consistent with our numerical results shown in Figure \ref{fig:tidal} for $\psi_\text{p} = 0^\circ$: larger values of $\tau$ cause the semi-major axis to increase on faster timescales than smaller values of $\tau$, and shorter planet rotation periods (larger $\Omega$) also lead to faster evolution of the semi-major axis. Therefore, we note that tides could potentially destabilize a moon with a prograde orbit if the planet has a very large $\tau$ value (small $Q$) or is rotating quickly, which could increase semi-major axis sufficiently after 3.1 Gyr to be beyond the empirically determined stability limit (see the blue dashed line in the right panel of Figure \ref{fig:tidal}). In fact, if $\tau$ is sufficiently large, any moons which formed close to the stability limit could migrate outward far enough to be stripped away from the planet by dynamical interactions. This scenario would require that $\tau$ be roughly two orders of magnitude larger than the value estimated for the Solar System giants. However, we cannot rule out this possibility because $\tau$ remains virtually unconstrained, especially for planets like \HIP, whose structure remains unknown with current measurements, and for which we have no close analogs in the Solar System.

Furthermore, we can gain insights regarding different moon scenarios from the scaling relation in Equation \ref{eq:timescale}. For example, this relation shows that less massive moons (which are presumably more common than the large moon we simulated here) migrate on a longer timescale, which means they would be less likely to be stripped away by dynamical interactions. Additionally, if \HIP~has a smaller radius than assumed here, then the migration timescale could drastically increase as well ($\tau_\text{mig} \propto R_\text{p}^{-5}$). This could be the case, for example, if \HIP~is indeed a smaller size planet surrounded by optically thick circumplanetary rings (although modeling such a scenario is beyond the scope of this work). This demonstrates that the cases we have chosen to simulate here probe near the upper limit of plausible moon configurations: decreasing the mass of the moon or shrinking the size of the planet would only serve to increase the chances of the moon's survival.

\subsection{Exomoon Observability}

Despite significant challenges in robustly detecting moons orbiting exoplanets, a number of efforts in recent years have attempted to uncover the first exomoon signals. Notably, the ``Hunt for Exomoons with \textit{Kepler}'' \citep[HEK;][]{Kipping+2012} systematically analyzed the light curves of a subset of all \textit{Kepler} planet candidates most amenable to hosting a moon. Using Bayesian multimodal nested sampling with a photodynamical forward model \citep[\texttt{LUNA};][]{Kipping+2011} to generate combined planet-moon transit light curves, \citet{Kipping+2013_II} constrained the satellite-to-planet mass ratios for seven potential satellite-hosting exoplanets, but did not find compelling evidence for an exomoon around any. Subsequent HEK searches also resulted in null exomoon detections in dozens of \textit{Kepler} systems \citep{Kipping+2015}, but set upper mass limits on satellites orbiting the habitable-zone planet Kepler-22b \citep{Kipping+2013_III} in addition to eight planets in M-dwarf systems \citep{Kipping+2014}. Finally, HEK has also constrained a tentative upper limit on the occurrence rate of Galilean-size moons around planets orbiting between 0.1 and 1 AU \citep{Teachey+2018}.

Although these surveys have not yet robustly confirmed any exomoons, they have detected at least two exomoon candidates. \citet{Teachey+2018_nature} reported evidence from transit observations and timing variations of the first exomoon candidate Kepler-1625b I, consistent with a Neptune-size satellite orbiting a Jovian-like planet. More recently, \citet{Kipping+2022} found tentative evidence supporting the existence of the exomoon candidate Kepler-1708b I, consistent with a 2.6 \Rearth~satellite. However, the veracity of either detection is the subject of ongoing discussion in the literature \citep[e.g.,][]{Kreidberg+2019, Teachey+2020, Cassese+2022}. 

Based on a suite of N-body and ET simulations in this work, we have shown that it is feasible for \HIP~to host a relatively large exomoon. Naturally, this leads to the question of whether such a moon could be detected with current and future facilities. While direct photodynamical searches for exomoons may be promising for other systems \citep[e.g., Kepler-1625b and Kepler-1708b;][]{Teachey+2018_nature, Kipping+2022}, several factors severely complicate such an effort for \HIP~given the existing data and the properties of the system. 

For example, unlike the original \textit{Kepler} mission, the extended K2 mission experienced significant systematic noise correlated with the spacecraft pointing \citep{Howell+2014}. Though it is possible to correct some correlated noise algorithmically \citep[e.g.,][]{Vanderburg+2016_k2sff}, residual systematics can persist at levels comparable to an exomoon signal, and stellar granulation and p-mode oscillations may persist as important noise sources. Moreover, for HST observations, systematic noise and gaps in the phase/time coverage introduced by the orbit of the telescope around Earth must be treated with caution, as these effects may also complicate detecting an exomoon in the light curve \citep[e.g.,][]{Kreidberg+2019}. For multiplanet systems such as HIP\,41378, one must also be wary of additional uncertainties in the inferred masses and orbits of other planets in the system, which can in turn lead to uncertain TTVs that affect the exomoon host (e.g., planet e). Without a firm grasp of the planet-planet-induced TTVs in this system, it is prohibitively difficult to disentangle potential TTVs caused by an orbiting moon, though in principle it may be possible for other systems \citep{Kipping_2021MNRAS}.

While these factors likely preclude a meaningful direct photodynamical search for exomoons orbiting \HIP~with current observations, it is still worth considering the effects that a large exomoon may have on transit observations. That is, although it is currently very difficult to detect exomoons (and robust confirmation is even more challening), they are likely included in extant and future observations of their exoplanet hosts. As our observations become more precise with newer technology (e.g., JWST and ELTs) it is necessary to consider exomoons as a source of uncertainty in our inferences about exoplanets, especially as we begin to probe the atmospheres of smaller planets in search of signatures of life. In this section we consider an idealized analog of \HIP~to explore how large exomoons can affect transit observations.

\subsubsection{White Light Curve}

\begin{figure*}[tbh]
    \centering
    \includegraphics[width=0.85\textwidth]{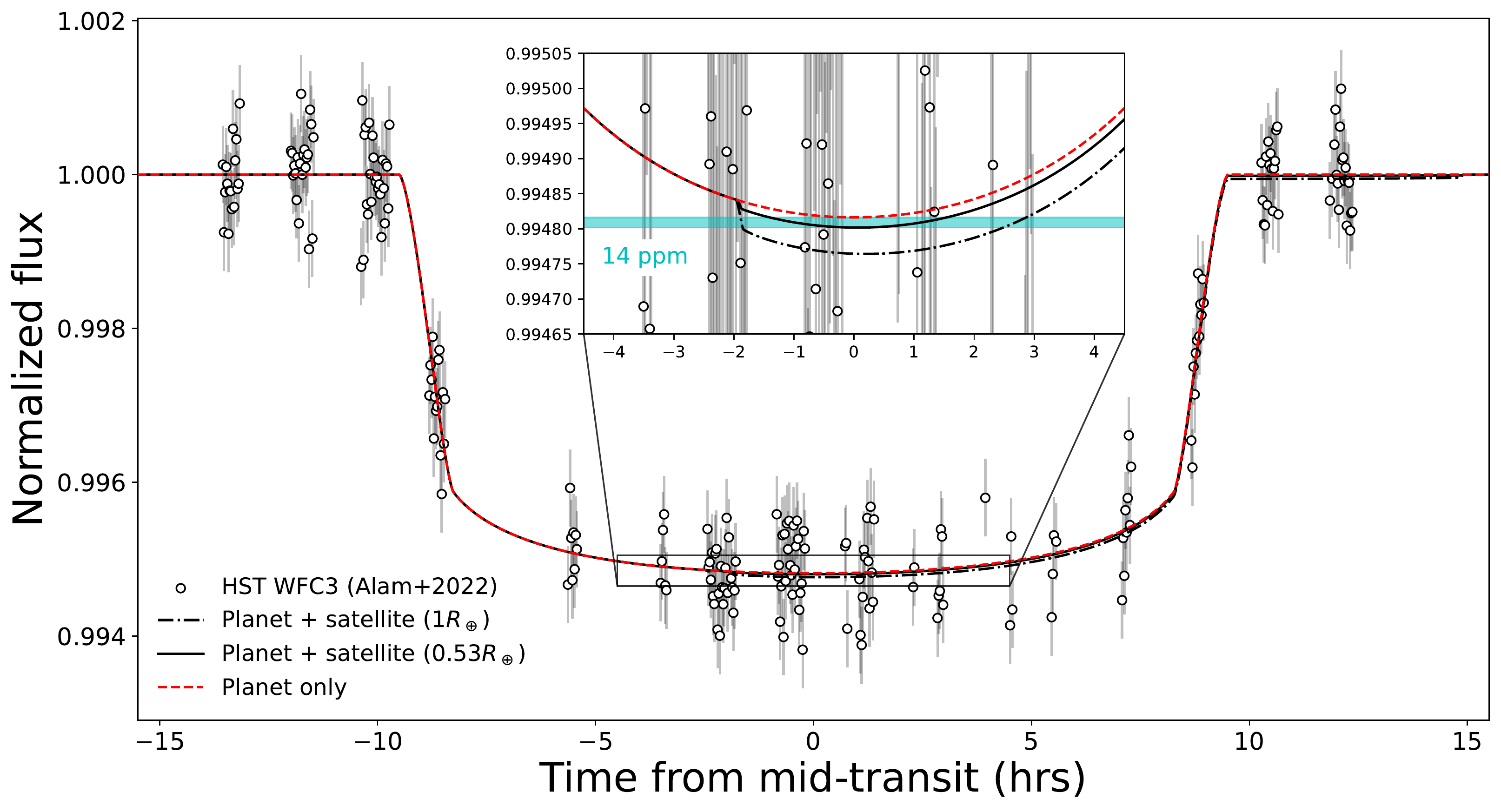}
    \caption{HST white light curve of \HIP~from \citet{Alam+2022} (points with errorbars) and best fit planet-only \texttt{Pandora} model (red dashed line). The black lines show simulated light curves with satellites injected at arbitrary orbital phase for ease of comparison---the solid line shows the signal from a 0.53 \Rearth~moon, while the dash-dotted line shows the signal from an Earth-size moon. Both simulated moons are assumed to have orbits aligned with the planet's reference plane and an orbital period of 30 days. A zoomed-in view near the transit center is shown in the inset panel, demonstrating the small effect produced by the presence of exomoons---the maximal flux difference between the moon-free and 0.53 \Rearth~moon models ($\sim$14 ppm; i.e., the transit depth of a 0.53 \Rearth~body transiting a 1.3 \Rsun~star) is highlighted in cyan. Posterior distributions for the planet-only model parameters are shown in Figure \ref{fig:corner}.}
    \label{fig:HST_lc}
\end{figure*}

To demonstrate the high signal-to-noise ratio required to detect a transiting exomoon, we modeled the fully detrended HST white light curve of \HIP~from \citet{Alam+2022} and injected the signal from a hypothetical moon orbiting the planet. We computed model light curves and evaluated the likelihood of each model using the \texttt{Pandora} package \citep{Hippke+2022}, an open-source photodynamical exomoon transit code optimized for computational speed and accuracy. First, for the planet-only model, we fixed the barycentric orbital period, semi-major axis, and inclination to the current literature values for \HIP~\citep{Santerne+2019}, and fit for the planet-to-star size ratio $R_p / R_\star$, quadratic limb darkening coefficients $q_1$ and $q_2$ \citep{Kipping_2013MNRAS_limbdarkening}, and an offset $\Delta T_0$ in the barycentric time of transit center $T_0$ to allow for small deviations from the transit center reported in \citet{Alam+2022}. We assumed a circular orbit for the planet. To effectively compute a planet-only model using \texttt{Pandora}, we tuned the satellite mass and radius to negligible values \citep[see][]{Hippke+2022}. 

Then, to derive posterior probability distributions for the planet-only fit, we used the nested sampling Monte Carlo algorithm \texttt{MLFriends} \citep{Buchner_2016, Buchner_2019} implemented in the \texttt{UltraNest}\footnote{\url{https://johannesbuchner.github.io/UltraNest/}} package \citep{Buchner_2021}. We imposed a Gaussian prior on $R_p / R_\star$ based on the results of \citet{Alam+2022}, and a conservative uniform prior on $\Delta T_0$ of $\pm0.1$ days\footnote{\citet{Alam+2022} reported a much smaller uncertainty in $T_0$ of about 0.002 days.}. For the limb-darkening coefficients, we estimated initial values with the \texttt{ExoTiC-LD} package \citep{hannah_wakeford_2022_6809899} using the 3D stellar models from \citet{Magic+2015} and stellar parameters from \citet{Santerne+2019}, then used these values to define Gaussian priors on $q_1$ and $q_2$ with a standard deviation of $0.1$. Using a Gaussian likelihood function, we then ran the \texttt{UltraNest} \texttt{ReactiveNestedSampler} with 800 live points, and a step sampler with 2000 steps, until convergence was achieved. We chose this particular nested sampling approach because it can be easily extended to conduct efficient retrievals of the full set of planet and moon parameters, whose posterior probability distributions can be multi-modal \citep[see][]{Hippke+2022}. Figure \ref{fig:HST_lc} shows the detrended white light curve from \citet{Alam+2022} along with the best-fit transit model from our nested sampling analysis. The full posterior probability distributions from the analysis are shown in Figure \ref{fig:corner} in the appendix. 

To investigate how a transiting exomoon would change the white light curve, we then calculated \texttt{Pandora} transit models with hypothetical 0.53 \Rearth~and 1 \Rearth~satellites injected on an arbitrary 30-day orbit ($a_\text{s} \approx 11\,R_\text{p} \approx 0.14\,R_\text{H}$) around \HIP. We calculated these models with \texttt{Pandora} assuming the best-fit parameters determined from the planet-only fit, but we set the moon radius to 0.53 \Rearth~or 1 \Rearth. In both cases we arbitrarily selected the moon's orbital phase for ease of comparison, and ignored the TTV caused by the moon by leaving the satellite mass set to be negligible. For comparison, these moon-injected transit models are shown along with the best-fit planet-only model in Figure \ref{fig:HST_lc}. The light curve models in Figure \ref{fig:HST_lc} clearly demonstrate the need for much higher precision ($\lesssim$15 ppm) than is available in current HST observations in order to photometrically detect exomoons around \HIP. 

We note that in an independent analysis of the HST/WFC3 transit of \HIP, \citet{Edwards+2022arXiv} obtained an RMS in the HST transit light curve of 125 ppm, compared to the 485 ppm RMS of \citet{Alam+2022}; since this precision is still $\gg$15 ppm, our conclusions here are independent of the choice of HST reduction. This high level of precision could, however, be within reach of JWST for certain systems---for example, \citet{Coulombe+2023arXiv} showed that the JWST NIRISS/SOSS white light curve of WASP-18b bins down to $\sim$5 ppm over one hour timescales, and \citet{Lustig-Yaeger+2023arXiv} demonstrated a similar result for LHS 475b using JWST NIRSpec/BOTS.

For completeness, we repeated the above procedure using observations from Campaigns 5 and 18 of the K2 mission \citep{Vanderburg+2016_hip, Santerne+2019}, which were detrended with \texttt{K2SFF} \citep{Vanderburg+2016_k2sff}. For each K2 light curve, we removed low-frequency variability by fitting a basis spline to the out-of-transit flux and dividing it out of the data \citep[as in][]{Berardo+2019}. We then repeated the procedure used to fit a planet-only model to the HST data, but we used the results from \citet{Santerne+2019} to define $T_0$ and set Gaussian priors on the $R_p / R_\star$, $q_1$, and $q_2$. The K2 data and best fit planet-only model are shown in Figure \ref{fig:K2_lc} in the appendix, along with the light curves with injected hypothetical satellites with radii of 0.53 \Rearth~and 1 \Rearth. The posterior probability distributions for the planet-only fit are shown in Figure \ref{fig:corner_k2} in the appendix. Similar to the HST observations, the K2 data are not precise enough to distinguish between planet-only and planet-moon transit models, and residual systematic variations at the $\gtrsim$20 ppm level can mimic the small signal produced by a moon.

Finally, we simulated an idealized single transit observation of \HIP~from the future PLAnetary Transits and Oscillation of stars (PLATO) mission, expected to launch in 2026 \citep{Rauer+2014}. We considered two scenarios: one in which only planet f transits a bright, photometrically quiet star with properties of HIP\,41378, and one in which the planet is orbited by an Earth-size satellite with the same configuration as previously described. We assumed a cadence of 600 s (10 min), as will be used for PLATO light curves, and a noise level of 20 ppm/hr for a $m_V\sim9$ star in PLATO's P1 sample \citep[though this number will actually depend on the number of cameras used to observe the target;][]{Rauer+2014}. Ignoring astrophysical variability, we added white noise to our simulated \texttt{Pandora} light curves that was randomly drawn from a Gaussian distribution with a standard deviation of 49 ppm (i.e., 20 ppm/hr scaled to 10-min cadence assuming Poisson noise). The transit models and simulated PLATO observations are shown in the bottom panel of Figure \ref{fig:K2_lc} in the appendix. Even in these idealized PLATO simulations, the photometric noise floor makes it extremely difficult to distinguish between the moon and moon-free scenarios (even in the absence of systematic uncertainties). Therefore, any variations on the order of 10 ppm due to instrumental and astrophysical systematics must be very well understood in order to robustly attempt to constrain the presence of an exomoon from transit observations alone.

\subsubsection{Implications for Transmission Spectroscopy}

\begin{figure*}[tbh]
    \centering
    \includegraphics[width=0.9\textwidth]{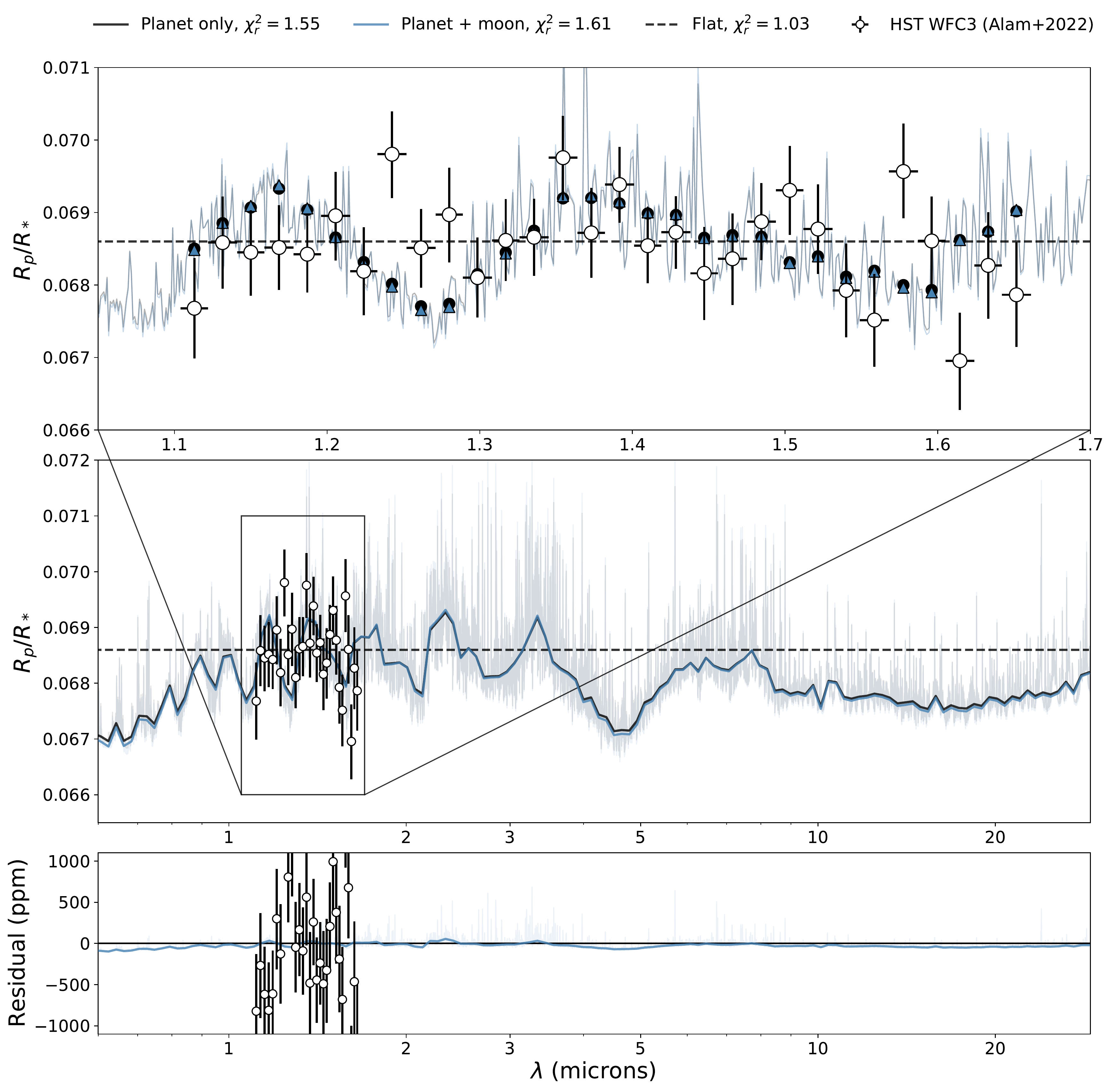}
    \caption{\textbf{Top:} Transmission spectrum of \HIP~\citep[points with error bars;][]{Alam+2022}, compared to 1D forward models and a flat line. For the planet-only model (black line), we assume a cloud-free 300x solar metallicity atmosphere and an isothermal ($T=294$ K) temperature profile; the combined planet and moon model (blue line) is described in the main text. The fainter lines show the computed unbinned ($R\sim1000$) \texttt{PLATON} spectra, while the colored points show each model binned to the same resolution as the WFC3 observations. \textbf{Middle:} \HIP~spectrum and models shown in the full \texttt{PLATON} wavelength range, including wavelengths that are accessible with JWST. The darker bold lines are binned by a factor of 30 to facilitate easier comparison. \textbf{Bottom:} Residual signal between the observed spectrum and the planet-only spectrum model, and between the planet-only spectrum model and the planet-moon model, over the same wavelength range.}
    \label{fig:spectra}
\end{figure*}

As observational capabilities continue to improve in the era of JWST and beyond, it is worth considering the extent to which exomoons may affect the observed wavelength dependence of transit depth, or transmission spectrum, of temperate and potentially habitable worlds. Current HST WFC3 observations of \HIP's atmosphere are consistent with a featureless transmission spectrum \citep[$\chi_r^2 \approx 1.05$;][]{Alam+2022}\footnote{We note that \citet{Edwards+2022arXiv} report a different transmission spectrum from their independent analysis of the HST/WFC3 data. We discuss this point later on in this section}. Given the low density of the planet, its flat spectrum could be explained by a variety of possible scenarios, including an extended hazy atmosphere, a high-metallicity clear atmosphere, or an optically thick ring system \citep{Alam+2022, Akinsanmi+2020}. Since these scenarios are all consistent with the observations, for simplicity we chose here to model \HIP~as if it had a high-metallicity clear atmosphere, and then assessed how an exomoon with a substantial atmosphere would alter its transmission spectrum\footnote{Note, however, that the high-metallicity clear scenario is probably the least likely atmospheric composition for this cool giant planet \citep{Alam+2022}, and it is possible that the ground truth is some combination of the scenarios we have described (e.g., the planet could have rings and a moon, rings and hazes, rings and hazes and a moon, etc.). While in principle each of these scenarios could affect the observed transmission spectrum of the system in a unique way, it is beyond the scope of this paper to model each possible ring/haze/moon permutation.}.

We computed model transmission spectra and evaluated the likelihood of each one using the \texttt{PLATON} package \citep{zhang+2019PASP_platon, Zhang+2020ApJ_platon}. First, we calculated the model for the planet based on the results of \citet{Alam+2022}, assuming a 300x solar metallicity atmosphere and an isothermal ($T=294$ K) T-P profile. Following \citet{Alam+2022}, we fit the model to the observed spectrum by binning the model predictions to the wavelength channels of the observations and performing a least-squares fit. Keeping all other parameters fixed, we fit for a constant vertical offset in $R_\text{p}/R_\star$ to preserve the shape of the spectrum. The observed spectrum and best fit 300x solar metallicity model are shown in Figure \ref{fig:spectra}.

We then computed a model for the combined transmission spectrum of a planet and a moon, assuming bodies both have substantial atmospheres and are in an optimal transit geometry (i.e., we see the maximum contribution of a potential moon's atmosphere). Following our stability simulations, we considered the scenario of a 0.53~\Rearth~(0.15 \Mearth) moon. For the hypothetical moon, we assumed the same isothermal T-P profile ($T=294$~K), but imposed a metallicity that would result in a maximally puffy atmosphere in order to explore an upper limit on the exomoon's signal in the transmission spectrum. 

We determined the lowest feasible atmospheric metallicity by considering the timescale associated with atmospheric boil-off due to an isothermal Parker wind \citep[see, for example,][]{Owen+2016ApJ, Wang+Dai_2019ApJ}. We set the boil-off timescale, $\tau_\text{boil}$, equal to the system age \citep[$\sim$3.1 Gyr;][]{Santerne+2019}, and solved for the mean molecular weight, $\mu$, using the following set of equations:
\begin{align}
    \tau_\text{boil} &\sim \frac{M_\text{atmo}}{\dot{M}_\text{Parker}}, \\
    \dot{M}_\text{Parker} &= 4 \pi r_\text{s}^2 c_\text{s} \rho_0 \exp\bigg( \frac{3}{2} - \frac{2 r_\text{s}}{R_\text{sat}} \bigg).
\end{align}
Here, $c_\text{s} = (k T / \mu)^{1/2}$ is the isothermal sound speed, where $k$ is the Boltzmann constant and $T$ is the isothermal temperature; $r_\text{s} = G M_\text{sat} / (2 c_\text{s}^2)$ is the sonic radius; and $\rho_0$ is the atmospheric mass density at the surface.

Assuming a surface pressure of 1 bar for the {0.53~\Rearth} moon's atmosphere (comparable to the atmospheric surface pressure on Titan), we found a minimum mean molecular weight of $\mu\sim4.1$, approximately 2x that of Jupiter and 1.5x that of Neptune. This corresponds to a metallicity of approximately 120x solar. We note, however, that this estimation is optimistic, as we do not consider other mass loss processes, such as photoevaporation. Importantly, we also note that we do not attempt to self-consistently model the moon's atmospheric or interior structure, nor how the moon acquired an atmosphere with this composition in the first place. The results presented here are simply intended as a thought experiment for a maximally-detectable exomoon, rather than a robust constraint on plausible exomoon atmospheres.

Under the assumption that the hypothetical moon would have had a viable formation pathway to achieve such a low-metallicity atmosphere, we computed its spectrum with \texttt{PLATON}, then combined this with the planet-only model as follows:
\begin{equation}
    (R_\text{p}/R_\star)_\text{eff} = \sqrt{\delta_\text{p} + \delta_\text{s}}
\end{equation}
where $(R_\text{p}/R_\star)_\text{eff}$ is the effective planet-to-star radius ratio due to the planet and moon, and $\delta_\text{p}$ and $\delta_\text{s}$ are the wavelength-dependent transit depths due to the planet and moon, respectively. Similar to before, we then performed a least-squares fit to the observed spectrum from \citet{Alam+2022}, preserving the shape of the total spectrum by fitting only for a constant offset in $R_\text{p}/R_\star$. Both models are shown alongside the observations in Figure \ref{fig:spectra}, along with the residual $R_\text{p}/R_\star$ between the planet-only and combined planet-moon models. 

Qualitatively, the addition of a moon with a substantial atmosphere results in a superposition of the moon and planet transmission spectra. Given our assumptions, the main differences between the planet-only spectrum and combined planet-moon spectrum are slightly enhanced H$_2$O absorption features ($<$50 ppm excess in $R_\text{p}/R_\star$) and a relative deficit in mid-IR and optical absorption ($<$100 ppm difference in $R_\text{p}/R_\star$) for the combined planet and moon models. We quantified the quality of each model fit, along with a flat line fit, using the reduced least-squares statistic. Like \citet{Alam+2022}, we cannot distinguish between the planet-only 300x solar metallicity model ($\chi_r^2 \approx 1.55$) or the flat line ($\chi_r^2 \approx 1.03$), given the data. Here we have also shown that the addition of the 0.53 \Rearth~exomoon with a $\mu\sim4.1$ atmosphere is indistinguishable compared to the planet-only model and the flat line ($\chi_r^2 \approx 1.61$). It is worth noting that the independent analysis of \citet{Edwards+2022arXiv} resulted in a different transmission spectrum from the HST/WFC3 data---their spectrum shows an upward slope from blue to red which is difficult to reproduce with equilibrium chemistry models and would lead to a qualitatively poorer fit to the forward models in Figure \ref{fig:spectra}. While we do not investigate the nature of the slope seen in \citet{Edwards+2022arXiv}, we conclude that using their spectrum would not change our inferences about the inability to detect an exomoon in the HST data.

Though the low-metallicity atmosphere we considered here is likely an extreme scenario and not necessarily likely for a moon orbiting \HIP, the excess absorption features in the model transmission spectrum may suggest that other systems containing large exomoons with thick atmospheres could experience non-negligible contamination if observed at high enough precision. This would be especially important for planets with smaller radii than we considered here. Such observations would need to be capable of resolving $\sim$10~ppm absorption features, a task that would be challenging even for JWST \citep[e.g.,][]{Lustig-Yaeger+2023arXiv, Coulombe+2023arXiv}.

For example, moons with substantial atmospheres could in principle contaminate otherwise high-fidelity spectra of terrestrial planets, introducing uncertainties in possible constraints on biosignature molecules. Moreover, the effects of a moon's atmosphere would vary in time, as the orbital phase of the moon at the time of observations would change with each transit. The temporal variability of the combined spectrum could therefore be misinterpreted as intrinsic temporal variability in the atmosphere of the planet. Finally, spectra of planets with large moons could be harder to interpret even if the moon itself lacks a substantial atmosphere. Akin to the transit light source effect \citep{Rackham+2018, Rackham+2019}, a moon could create an apparent time-varying inhomogeneity on the disk of the host star, leading to false spectral features. While the scenarios discussed thus far are only for singular large moons, we note that the presence of multiple smaller moons (like we see in the outer Solar System) could also introduce additional timescales and complexity to the problem. Therefore, it may be important to consider moons as a source of uncertainty in transmission spectroscopy as advances in technology allow for more precise measurements of exoplanet atmospheres.

\section{Conclusion} \label{sec:conclusion}

In this work, we investigated the stability of a hypothetical large moon (0.15 \Mearth, 0.53 \Rearth) orbiting the long-period ($P \approx 1.5$ yr) 12 \Mearth~planet \HIP. Using a suite of numerical N-body simulations, we have shown that moons up to this size can survive in the system over long timescales over a wide range of system setups, including with the added complexity of multiple planets in the HIP\,41378 system and tidal interactions between the moon and the host planet. Although our analysis has demonstrated the plausibility of a moon orbiting \HIP, we have shown that the detection of a relatively large Earth-size moon in the system is unlikely given current observations, but may be feasible in the near future with high-precision JWST observations. Moreover, an exomoon with a sufficiently puffy atmosphere, while highly idealized, could imprint $\sim$10 ppm features on the transmission spectrum of the planet (though it is unclear whether this signal could be recognized as such). Our main conclusions are summarized as follows:
\begin{enumerate}
    \item \HIP~could feasibly host exomoons at least as massive as 0.15 \Mearth~that are stable over a timescale of $10^5$ yr, in agreement with stability limits determined by previous works \citep[e.g.,][]{Rosario-Franco+2020AJ}. Under ideal assumptions of zero eccentricity and co-planar orbits, a 0.15 \Mearth~moon does not escape planet f's gravitational sphere of influence or collide with the planet as long as it orbits within $\sim$40\% of the planet's Hill radius. Our simulations demonstrate that this limit does not depend strongly on $\cos i_\text{s}$ for most (low) moon inclinations. However, Kozai-Lidov oscillations can start to become an important destabilizing influence for $\cos i_\text{s} \gtrsim 0.8$, and disallow stable orbits for the highest inclinations ($\cos i_\text{s} \gtrsim 0.95$). The satellite mass chosen here represents an extreme scenario, analogous to the highest moon-to-planet mass ratio observed in the Solar System.
    \item Current constraints on the mass and orbit of the inner planet HIP\,41378\,\,e do not necessarily preclude the possibility of a moon orbiting planet f. However, improved measurements of both planets' masses and orbits are needed to robustly constrain the stability of a putative exomoon in the system. If, for example, the true eccentricity of planet e is 1$\sigma$ smaller than the eccentricity reported by \citet{Santerne+2019}, then moons orbiting within $\sim$40\% of the Hill radius of planet f should be stable, as long as $0 < \cos i_\text{s} \lesssim 0.9$. This not only highlights the need for improved masses and orbital constraints for the planets in the HIP 41378 system, but also the importance of considering multiple planets in dynamical modeling of exomoon stability in multi-planet systems in general.
    \item For most circular prograde orbits, tidal interactions between planet f and a moon are not sufficient to cause the moon to migrate outward beyond the 0.4$R_H$ dynamical stability limit, assuming that planet f has tidal response properties comparable to the Solar System giant planets. However, under extreme circumstances where planet f is given a very large time lag ($\tau \gtrsim 0.7$ s) and fast rotation rate ($P\sim3$ hr), tidal migration can lead to the moon's escape from the gravitational influence of the planet. While in principle more massive moons would also migrate faster, we do not consider such cases here because moons with higher mass ratios have not been confirmed in the Solar System or in extrasolar systems. For high planet obliquities (or moon inclinations), tidal interactions alone are not sufficient to destabilize the moon, but, as mentioned in the first point, Kozai-Lidov oscillations can become significant enough to destabilize such systems. We also note that moons with retrograde orbits are more likely to collide with the planet due to inward tidal migration (independent of the dynamical stability limit), but this depends strongly on the assumed tidal parameters of the planet, which remain unconstrained.
    \item Existing observations of the HIP\,41378 system from HST and K2 cannot reliably constrain the presence of exomoons, as the expected transit signal produced by a large moon (0.53 \Rearth) is only of the order 15ppm. Furthermore, uncertainties in the orbits and masses of the planets in the HIP\,41378 system, as well as correspondingly uncertain TTVs, likely preclude the detection of a moon with planned missions such as PLATO. However, it is possible that future observations with JWST capable of precision better than $\sim$20~ppm on 10-minute timescales will be able to detect exomoons in other systems with more robust planet measurements. For example, the giant temperate planet PH-2b (Kepler-86b) is currently slated to be observed by JWST NIRSpec/PRISM in 2024 (GO Cycle 2 Proposal ID: 3235; PI: Fortney), and could potentially provide serendipitous observations of a transiting exomoon. 
    \item Exomoons with thick atmospheres may contaminate the transmission spectra of exoplanets at the $\sim$10 ppm level in a time-dependent manner (well below the precision of HST observations). However, this is an optimistic estimate which assumes that a moon can acquire and retain a low mean molecular weight atmosphere over its lifetime. Nonetheless, exomoons with significant atmospheres may be sources of uncertainty in other exoplanet systems as new technology enables smaller and smaller signals to be probed.
\end{enumerate}

We note that while our analysis provides conservative limits on the stability of exomoons orbiting \HIP~using well-tested methods such as N-body simulations and equilibrium tide modeling, there are a few caveats which may be important to consider in the real HIP\,41378 system. For example, multiple planets in the system do not have well-constrained orbital periods or masses due to the difficulty of measuring long-period planets ($P\gtrsim1$ yr). While we attempted to incorporate some of this uncertainty in our analysis (e.g., randomly drawing masses and orbital parameters for planet e in our N-body simulations), we could not account for all the unconstrained degrees of freedom in the actual HIP\,41378 system, which includes at least 5 planets in total. Moreover, we did not attempt to model the effects of multiple moons in the system, which also would have increased the complexity of our simulations.

Given the current data, another caveat is that we do not know with certainty what the true nature of \HIP~is. Throughout our analysis we assumed that the planet has a mass and radius consistent with observations \citep{Santerne+2019} and tidal properties similar to a Jovian planet. However, if in reality the planet is more similar to Neptune and it possess a circumplanetary ring system \citep[e.g.,][]{Akinsanmi+2020, Piro+Vissapragada_2020}, this could change the tidal response of the planet and therefore affect the orbital stability of moons. 

Moreover, self-consistent modelling of any moon-ring interactions is beyond the scope of this work, but would be an interesting subject of future simulations. \citet{Akinsanmi+2020} showed that observations of \HIP~are consistent with a planet of radius $R_\text{p} = 3.7$ \Rearth~($\rho_\text{p} = 1.4 \,\text{g cm}^{-3}$), with opaque rings extending from 1.05 to 2.6 \Rearth. If the rings disperse out to the Roche radius, it follows that the density of the ring material would be $\rho_\text{r} = 1.08 \,\text{g cm}^{-3}$. In this case, assuming that the ring particles have a typical radius of about 100 cm, the total mass of the rings would be approximately $10^{19}$ kg \citep[see Equation 4 of][]{Piro_2018_rings}, roughly the observed mass of Saturn's rings \citep{Iess+2019}. Hence in reality, gravitational interactions between the rings and a moon may affect their long-term stability \citep[e.g.,][]{Nakajima+2020}. On a similar note, we did not attempt to model the combined observable effects of moons and rings \citep[e.g.,][]{Ohno+2022}.

Finally, throughout our analysis we assumed that moons are a natural outcome of planet formation, and we only considered the stability of moons once they reached a tidally-locked state with planet f. That is, we did not attempt to treat exomoon formation and evolution in a self-consistent manner, and we remained agnostic to the specific formation pathway of our simulated moons.

To assess the observability of an exomoon, we chose to model transit observations because these are likely the most promising for a long-period transiting planet like \HIP. We note that aside from photodynamical transit searches for exomoons such as HEK \citep{Kipping+2012} and the future Transiting Exosatellites, Moons, and Planets in Orion (TEMPO) survey \citep{Limbach+2022}, several other methods of detecting exomoons have been proposed. For example, the orbital sampling effect (OSE) technique \citep{Heller+2014_ose}, which leverages many transit observations to achieve robust statistics, can be used to infer the size, orbital separation, and mass of transiting exomoons from a phase-folded light curve of about a dozen transits \citep{Hippke_2015, Heller+2016}. However, this is not currently feasible for \HIP~because the planet's long orbital period and transit duration make it difficult to acquire the necessary large number of observations.

Other novel approaches, including direct imaging and Doppler spectroscopy \citep{Peters+Turner_2013, Agol+2015, Vanderburg+2018}, radio emission from satellite-hosting exoplanets \citep{Noyola+2014, Noyola+2016, Narang+2022arXiv}, and combined planet-moon thermal phase curves \citep{Forgan_2017}, may also be fruitful probes of exomoons with future technologies. But again, \HIP~is not an optimal target for such studies. The semi-major axis of planet f is too small for current direct imaging technology, with a sky-projected angular separation of $\sim$13~mas, and uncertainties in the other planet masses and orbits preclude a radial velocity exomoon search. Moreover, uncertainty regarding the nature of planet f make radio emission studies challenging, and the cool equilibrium temperature of planet f ($T_\text{eq} = 294$~K) is not sufficient to produce an appreciable thermal phase curve.

Nonetheless, as our ability to measure exoplanets continues to improve in the future, exomoons may indeed become a more important source of uncertainty (in addition to transit light source variability, stellar spectrum noise, etc.). Further into the future, we may even be able to robustly confirm detections of exomoons and constrain the properties of their atmospheres with state-of-the-art observational facilities such as ground-based extremely large telescopes (ELTs) and next-generation space missions like the Habitable Worlds Observatory (HWO). For example, the Ancillary Science Case A-19 for the Large UV/Optical/Infrared Surveyor (LUVOIR) suggests spectroastrometry (i.e., measuring the astrometric shift that occurs between wavelengths with flux dominated by the exoplanet and wavelengths dominated by the exomoon) as a possible exomoon detection strategy with a 12~m class space telescope \citep{LUVOIR_2019arXiv}. With contrast $>10^{-9}$ and relative astrometric precision between wavelength bands of $\sim$0.1 mas, such an observatory could in principle detect an Earth-size exomoon orbiting a 1~AU Jupiter-size planet at 10~pc \citep{LUVOIR_2019arXiv}. Not only would this be a revolutionary discovery in itself, but it would also open the possibility of using exomoons as a way of studying planet formation and evolution, and even provide a potential new avenue for searching for signs of life beyond the Solar System.

\bigskip
The authors thank Kristo Ment for helpful discussions regarding quadtree algorithms.
This paper makes use of observations from the NASA/ESA Hubble Space Telescope, obtained at the Space Telescope Science Institute, which is operated by the Association of Universities for Research in Astronomy, Inc., under NASA contract NAS 5–26555. 
These observations are associated with HST-GO program 16267 (PI: Dressing), and the analysis was supported by grant HST-GO-16267. This work was supported by the Programme National de Plan\'etologie (PNP) of CNRS-INSU co-funded by CNES. 
This work was supported by FCT - Funda\c{c}\~{a}o para a Ci\^{e}ncia e a Tecnologia through national funds and by FEDER through COMPETE2020 - Programa Operacional Competitividade e Internacionaliza\c{c}\~{a}o by these grants: UIDB/04434/2020; UIDP/04434/2020, 2022.06962.PTDC. 
This research has also been partly funded by the Spanish State Research Agency (AEI) Project No.~PID2019-107061GB-C61.
ACC~acknowledges support from STFC consolidated grant numbers ST/R000824/1 and ST/V000861/1, and UKSA grant number ST/R003203/1.
CKH~acknowledges support from the National Science Foundation Graduate Research Fellowship Program under Grant No.~DGE 2146752.
JK~acknowledges financial support from Imperial College London through an Imperial College Research Fellowship grant.
JL-B~acknowledges support from the Ram\'on y Cajal programme (RYC2021-031640-I) supported by the MCIN/AEI/10.13039/501100011033 and the European Union ``NextGenerationEU''/PRTR as well as partial financial support received from ``la Caixa'' Foundation (ID 100010434) and the European Unions Horizon 2020 research and innovation programme No 847648, with fellowship code LCF/BQ/PI20/11760023. 
NCS~acknowledges funding by the European Union (ERC, FIERCE, 101052347). Views and opinions expressed are however those of the author(s) only and do not necessarily reflect those of the European Union or the European Research Council. Neither the European Union nor the granting authority can be held responsible for them.
SGS~acknowledges the support from FCT through Investigador FCT contract nr.~CEECIND/00826/2018 and POPH/FSE (EC).
PJW~acknowledges support from the UK Science and Technology Facilities Council (STFC) under consolidated grant ST/T000406/1.


\vspace{5mm}
\facilities{HST (WFC3), K2}

\software{
Astropy \citep{astropy_2013, astropy_2018},
AstroQTpy \citep{harada_astroQTpy_2023},
EqTide \citep{Barnes_2017},
ExoTiC-LD \citep{hannah_wakeford_2022_6809899},
IPython \citep{Perez+2007},
Matplotlib \citep{Hunter+2007},
NumPy \citep{vanderWalt+2011, Harris+2020},
Pandora \citep{Hippke+2022},
PLATON \citep{zhang+2019PASP_platon,Zhang+2020ApJ_platon},
Rebound \citep{rebound, reboundias15},
UltraNest \citep{Buchner_2021}
}

\appendix

\section{Benchmark three-body simulation}\label{app:three-body}

While in the main text we describe our N-body simulations for exploring how the semi-major axis and inclination of the satellite and the presence of planet e affect the system's stability, here we describe the set of intial simulations that were used to test our N-body code.

We benchmarked our code by following the general procedure outlined by \citet{Domingos+2006} and \citet{Rosario-Franco+2020AJ}. We set up our grid of simulations using a quadtree data structure implemented in \texttt{astrQTpy} \citep{harada_astroQTpy_2023}, which we described in more detail in Section \ref{sec:methods:nbody:threebody}. We constrained the initial eccentricity, $e_\text{f}$, of planet f between 0.0 and 0.5, and the initial semi-major axis, $a_\text{s}$, of the satellite between 0.25 to 0.55$R_\text{H}$, where $R_\text{H}$ is the Hill radius defined in Equation \ref{eq:hill} ($R_\text{H} \approx 0.03$ AU $\approx 76 R_\text{p}$). These limits were chosen in order to explore the parameter space encompassing the stability limit determined by \citet{Rosario-Franco+2020AJ}. Note however that the measured eccentricity of planet f is actually constrained to $e_\text{f} = 0.004_{-0.003}^{+0.009}$ \citep{Santerne+2019}, as stated in the main text. As in Section \ref{sec:methods:nbody:threebody}, the planet and satellite were initialized with co-planar orbits, with the longitudes of the ascending node and arguments of pericenter set to zero ($\Omega = \omega = 0^\circ$), and the mean anomaly of the planet set to zero ($\mathcal{M}_\text{f} = 0^\circ$).

\begin{figure*}[bt!]
    \centering
    \includegraphics[width=0.8\textwidth]{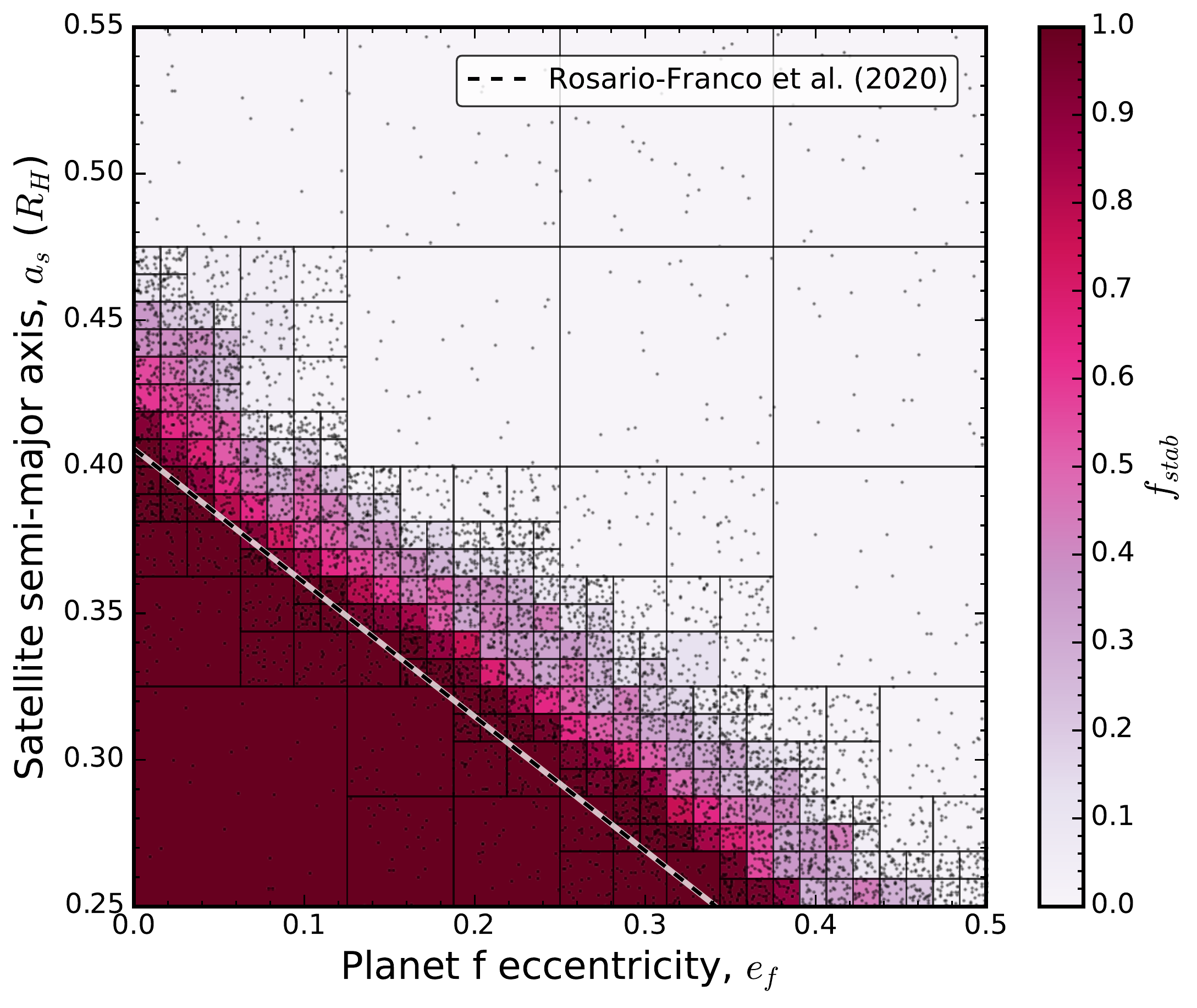}
    \caption{Three-body orbital stability map for a system consisting of HIP\,41378, planet f, and planet f's satellite, as a function of planetary eccentricity, $e_\text{f}$, and satellite semi-major, axis $a_\text{s}$ (expressed in terms of the Hill radius, $R_\text{H}$). The color of each grid cell corresponds to the fraction $f_\text{stab}$ of 25 simulations within each cell (indicated by the black points) that survive over $10^5$ yr. The stability limit determined by \citet{Rosario-Franco+2020AJ}, shown as the dashed line, indicated good agreement with our simulations at low eccentricities.}
    \label{fig:nbody-a_moon-vs-ecc_f}
\end{figure*}

We proceeded by dividing the parameter space evenly into a $4 \times 4$ grid, and running 25 independent N-body simulations in each grid cell with $(e_\text{f}, a_\text{s})$ and the satellite mean anomaly, $\mathcal{M}_\text{s}$, drawn from uniform distributions. We calculated the fraction of stable systems, $f_\text{stab}$, in each cell as the fraction of simulations (out of 25) which satisfied the stability criteria defined in the main text over a timescale of $10^5$ yr. We then continued splitting each grid cell in the quadtree into four equal child cells and computing more N-body simulations, following the procedure described in Section \ref{sec:methods:nbody:threebody}. Figure \ref{fig:nbody-a_moon-vs-ecc_f} shows the final stability grid generated by our simulations, where $f_\text{stab}$ is quantified by color. Regions where $0 < f_\text{stab} < 1$ are considered quasi-stable, while regions where $f_\text{stab} = 1$ (0) are absolutely stable (unstable) over $10^5$ yr.

Our results were qualitatively similar to \citet{Rosario-Franco+2020AJ}, and quantitatively consistent in terms of the stability boundary for low eccentricity orbits (i.e., the largest value of $a_\text{s}$ for a given $e_\text{f}$ where $f_\text{stab} = 1$). \citet{Rosario-Franco+2020AJ} modeled the stability boundary as 
\begin{equation}\label{eq:stability_limit}
    a_\text{crit} = c_1 (1 - c_2 e_f)
\end{equation}
where $a_\text{crit}$ is implicitly in terms of the fraction of the Hill radius, $c_1$ is the stability limit for circular orbits, and $c_2$ is a slope parameter.

From their N-body simulations, the authors determined that $c_1 = 0.4061 \pm 0.0028$ and $c_2 = 1.1257 \pm 0.0273$. Their stability limit is shown in Figure \ref{fig:nbody-a_moon-vs-ecc_f} plotted over our stability. Our results are in good agreement for low eccentricities, with our stability limit for circular orbits being $a_\text{crit} \approx 0.41 R_\text{H}$. However, we note that our simulations tend to predict a slightly higher stability limit as eccentricity increases (for $e_\text{f} \gtrsim 0.15$). This discrepancy at higher eccentricities is likely due to our different choice of planetary semi-major axis; we assumed a semi-major axis of 1.37 AU for \HIP, whereas \citet{Rosario-Franco+2020AJ} assumed an orbital separation of 1 AU in their simulations. Because our simulated planet and its moon are farther away from the host star, the system tends to be more stable against gravitational perturbations.

\section{Additional Figures}\label{app:figures}

\begin{figure*}[bth]
    \centering
    \includegraphics[width=0.85\textwidth]{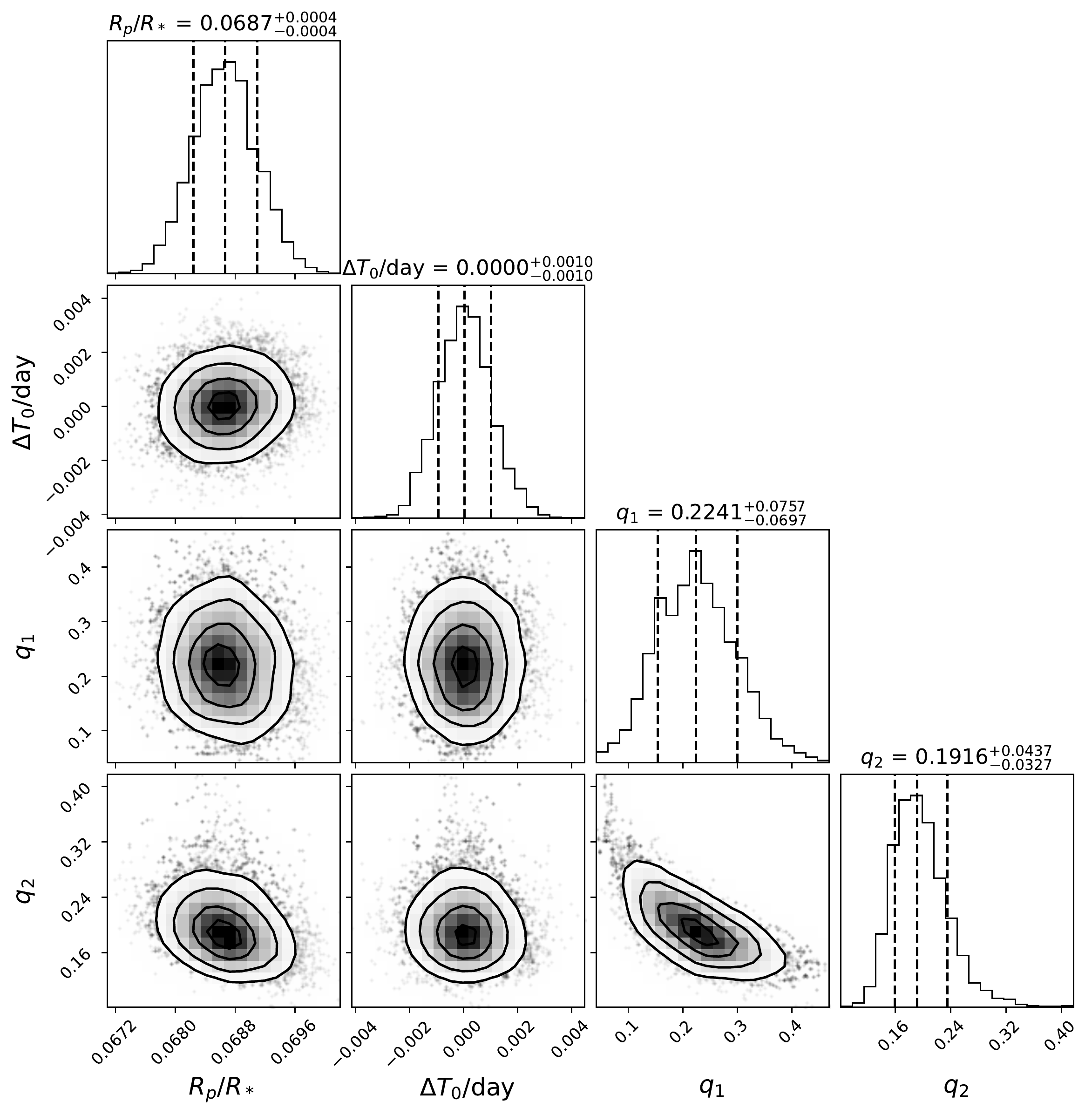}
    \caption{HST white light curve posterior distributions for the planet-only model shown in Figure \ref{fig:HST_lc}.}
    \label{fig:corner}
\end{figure*}

\begin{figure*}[bth]
    \centering
    \includegraphics[width=0.85\textwidth]{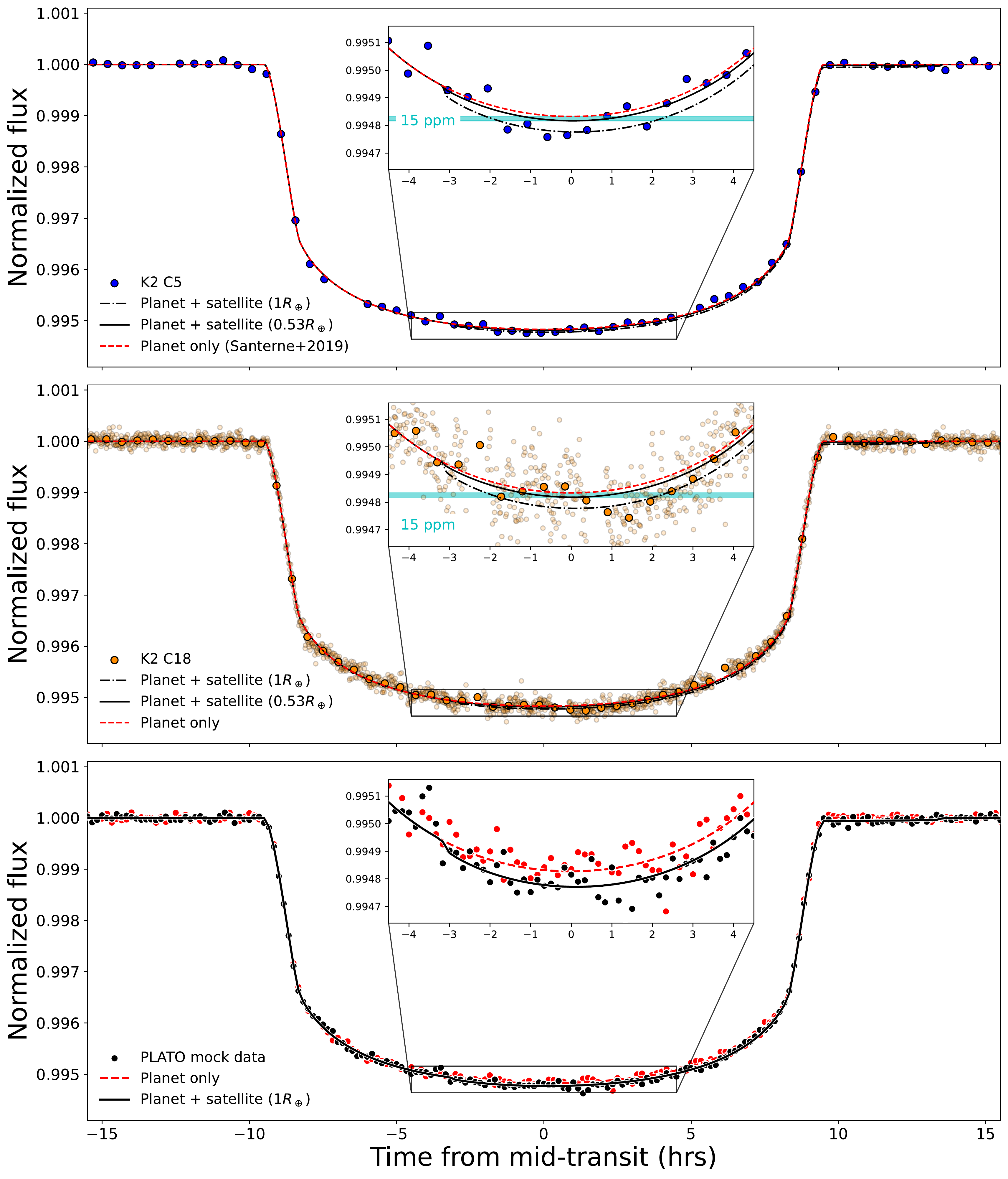}
    \caption{\textbf{Top:} K2 transit light curve for \HIP~from Campaigns 5 (blue points) with best fit planet-only model (red dashed line). As in Figure \ref{fig:HST_lc}, additional models with injected signals from a 0.53 \Rearth~moon and an Earth-size moon are shown by the black solid line and dash-dotted line. The inset panel shows a zoomed-in view near mid-transit, demonstrating that the effects of moons are below the noise level of the data. Note that the 15 ppm difference in mid-transit flux between the moon-free and 0.53 \Rearth~moon cases is slightly greater in the K2 bandpass than the HST bandpass (Figure \ref{fig:HST_lc}) due to differences in the stellar limb-darkening coefficients. \textbf{Middle:} Same as the top panel, but with data from K2 Campaign 18. Short-cadence data are shown by the transparent points, while the opaque points have been binned to match the long-cadence data. \textbf{Bottom:} Simulated 10-min cadence PLATO data for a single transit of \HIP~assuming the presence of no moon (red points) and an Earth-size moon (black points). The underlying \texttt{Pandora} models for each case are shown by the lines with corresponding colors, and a zoomed-in view near mid-transit is shown in the inset panel.}
    \label{fig:K2_lc}
\end{figure*}

\begin{figure*}[bth]
    \centering
    \includegraphics[width=0.95\textwidth]{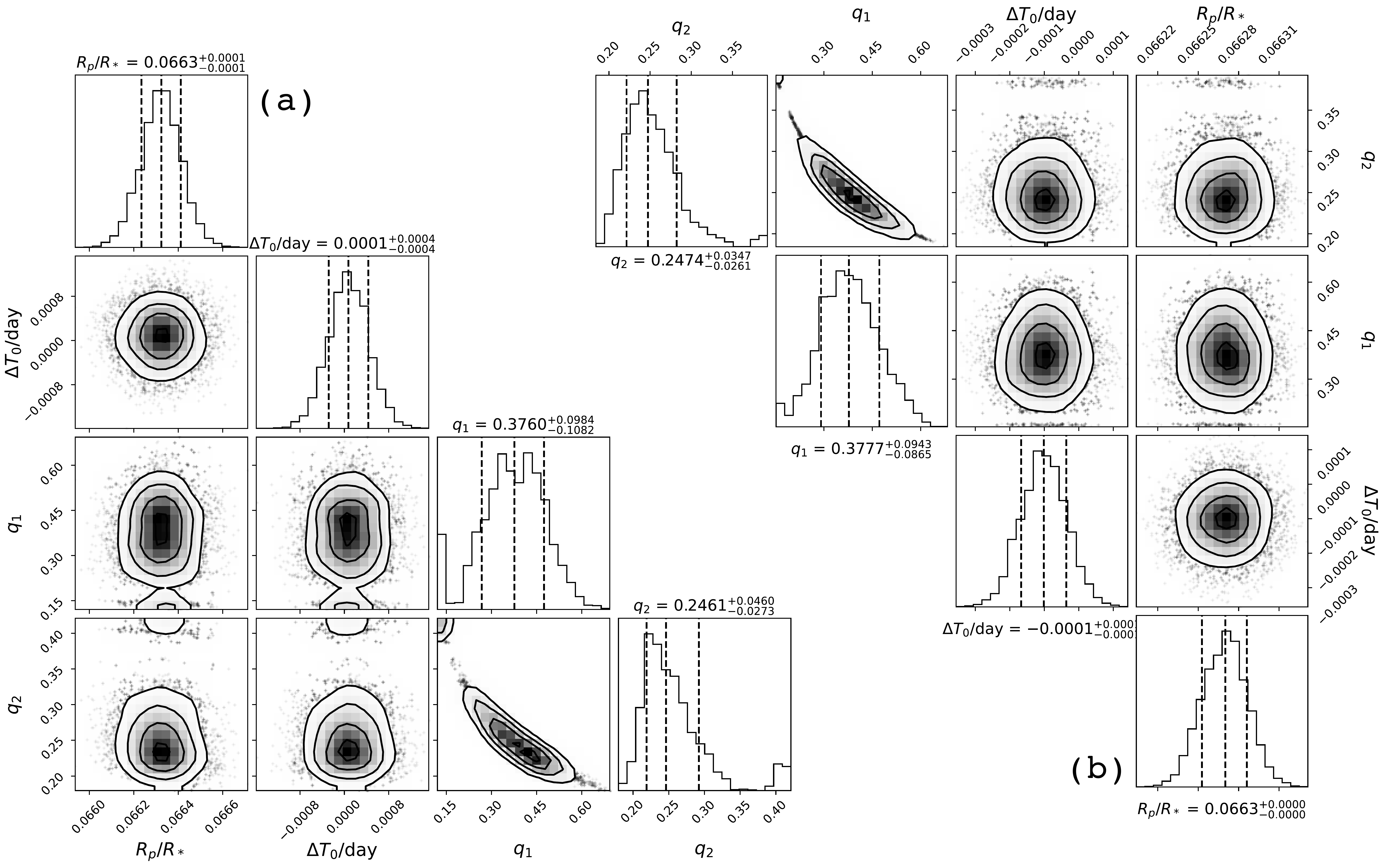}
    \caption{(a) K2 C5 light curve posterior distributions for the planet-only model shown in the top panel of Figure \ref{fig:K2_lc}. (b) K2 C18 light curve posterior distributions for the planet-only model shown in the middle panel of Figure \ref{fig:K2_lc}.}
    \label{fig:corner_k2}
\end{figure*}

Here we present additional figures that supplement the main text. Figure \ref{fig:corner} shows the posterior distributions derived from our nested sampling analysis of the HST white light curve shown in Figure \ref{fig:HST_lc}---maximum likelihood values from the posteriors were used to compute the planet-only model therein. Figure \ref{fig:K2_lc} shows additional observations of \HIP~from K2 Campaigns 5 and 18, along with simulated data of a single transit event from the upcoming PLATO mission. The posterior distributions derived from nested sampling analyses of the K2 light curves are shown in Figure \ref{fig:corner_k2}, which were used to calculate the maximum likelihood planet-only transit models.

\bibliography{main}{}

\begin{thebibliography}{}
\expandafter\ifx\csname natexlab\endcsname\relax\def\natexlab#1{#1}\fi
\providecommand{\url}[1]{\href{#1}{#1}}
\providecommand{\dodoi}[1]{doi:~\href{http://doi.org/#1}{\nolinkurl{#1}}}
\providecommand{\doeprint}[1]{\href{http://ascl.net/#1}{\nolinkurl{http://ascl.net/#1}}}
\providecommand{\doarXiv}[1]{\href{https://arxiv.org/abs/#1}{\nolinkurl{https://arxiv.org/abs/#1}}}

\bibitem[{{Agol} {et~al.}(2015){Agol}, {Jansen}, {Lacy}, {Robinson}, \&
  {Meadows}}]{Agol+2015}
{Agol}, E., {Jansen}, T., {Lacy}, B., {Robinson}, T.~D., \& {Meadows}, V. 2015,
  \apj, 812, 5, \dodoi{10.1088/0004-637X/812/1/5}

\bibitem[{{Akinsanmi} {et~al.}(2019){Akinsanmi}, {Barros}, {Santos}, {Correia},
  {Maxted}, {Bou{\'e}}, \& {Laskar}}]{Akinsanmi+2019A&A}
{Akinsanmi}, B., {Barros}, S.~C.~C., {Santos}, N.~C., {et~al.} 2019, \aap, 621,
  A117, \dodoi{10.1051/0004-6361/201834215}

\bibitem[{{Akinsanmi} {et~al.}(2020){Akinsanmi}, {Santos}, {Faria}, {Oshagh},
  {Barros}, {Santerne}, \& {Charnoz}}]{Akinsanmi+2020}
{Akinsanmi}, B., {Santos}, N.~C., {Faria}, J.~P., {et~al.} 2020, \aap, 635, L8,
  \dodoi{10.1051/0004-6361/202037618}

\bibitem[{{Alam} {et~al.}(2022){Alam}, {Kirk}, {Dressing}, {L{\'o}pez-Morales},
  {Ohno}, {Gao}, {Akinsanmi}, {Santerne}, {Grouffal}, {Adibekyan}, {Barros},
  {Buchhave}, {Crossfield}, {Dai}, {Deleuil}, {Giacalone}, {Lillo-Box},
  {Marley}, {Mayo}, {Mortier}, {Santos}, {Sousa}, {Turtelboom}, {Wheatley}, \&
  {Vanderburg}}]{Alam+2022}
{Alam}, M.~K., {Kirk}, J., {Dressing}, C.~D., {et~al.} 2022, \apjl, 927, L5,
  \dodoi{10.3847/2041-8213/ac559d}

\bibitem[{{Alvarado-Montes} {et~al.}(2017){Alvarado-Montes}, {Zuluaga}, \&
  {Sucerquia}}]{Alvarado-Montes+2017}
{Alvarado-Montes}, J.~A., {Zuluaga}, J.~I., \& {Sucerquia}, M. 2017, \mnras,
  471, 3019, \dodoi{10.1093/mnras/stx1745}

\bibitem[{{Astropy Collaboration} {et~al.}(2013){Astropy Collaboration},
  {Robitaille}, {Tollerud}, {Greenfield}, {Droettboom}, {Bray}, {Aldcroft},
  {Davis}, {Ginsburg}, {Price-Whelan}, {Kerzendorf}, {Conley}, {Crighton},
  {Barbary}, {Muna}, {Ferguson}, {Grollier}, {Parikh}, {Nair}, {Unther},
  {Deil}, {Woillez}, {Conseil}, {Kramer}, {Turner}, {Singer}, {Fox}, {Weaver},
  {Zabalza}, {Edwards}, {Azalee Bostroem}, {Burke}, {Casey}, {Crawford},
  {Dencheva}, {Ely}, {Jenness}, {Labrie}, {Lim}, {Pierfederici}, {Pontzen},
  {Ptak}, {Refsdal}, {Servillat}, \& {Streicher}}]{astropy_2013}
{Astropy Collaboration}, {Robitaille}, T.~P., {Tollerud}, E.~J., {et~al.} 2013,
  \aap, 558, A33, \dodoi{10.1051/0004-6361/201322068}

\bibitem[{{Astropy Collaboration} {et~al.}(2018){Astropy Collaboration},
  {Price-Whelan}, {Sip{\H{o}}cz}, {G{\"u}nther}, {Lim}, {Crawford}, {Conseil},
  {Shupe}, {Craig}, {Dencheva}, {Ginsburg}, {VanderPlas}, {Bradley},
  {P{\'e}rez-Su{\'a}rez}, {de Val-Borro}, {Aldcroft}, {Cruz}, {Robitaille},
  {Tollerud}, {Ardelean}, {Babej}, {Bach}, {Bachetti}, {Bakanov}, {Bamford},
  {Barentsen}, {Barmby}, {Baumbach}, {Berry}, {Biscani}, {Boquien}, {Bostroem},
  {Bouma}, {Brammer}, {Bray}, {Breytenbach}, {Buddelmeijer}, {Burke},
  {Calderone}, {Cano Rodr{\'\i}guez}, {Cara}, {Cardoso}, {Cheedella}, {Copin},
  {Corrales}, {Crichton}, {D'Avella}, {Deil}, {Depagne}, {Dietrich}, {Donath},
  {Droettboom}, {Earl}, {Erben}, {Fabbro}, {Ferreira}, {Finethy}, {Fox},
  {Garrison}, {Gibbons}, {Goldstein}, {Gommers}, {Greco}, {Greenfield},
  {Groener}, {Grollier}, {Hagen}, {Hirst}, {Homeier}, {Horton}, {Hosseinzadeh},
  {Hu}, {Hunkeler}, {Ivezi{\'c}}, {Jain}, {Jenness}, {Kanarek}, {Kendrew},
  {Kern}, {Kerzendorf}, {Khvalko}, {King}, {Kirkby}, {Kulkarni}, {Kumar},
  {Lee}, {Lenz}, {Littlefair}, {Ma}, {Macleod}, {Mastropietro}, {McCully},
  {Montagnac}, {Morris}, {Mueller}, {Mumford}, {Muna}, {Murphy}, {Nelson},
  {Nguyen}, {Ninan}, {N{\"o}the}, {Ogaz}, {Oh}, {Parejko}, {Parley}, {Pascual},
  {Patil}, {Patil}, {Plunkett}, {Prochaska}, {Rastogi}, {Reddy Janga},
  {Sabater}, {Sakurikar}, {Seifert}, {Sherbert}, {Sherwood-Taylor}, {Shih},
  {Sick}, {Silbiger}, {Singanamalla}, {Singer}, {Sladen}, {Sooley},
  {Sornarajah}, {Streicher}, {Teuben}, {Thomas}, {Tremblay}, {Turner},
  {Terr{\'o}n}, {van Kerkwijk}, {de la Vega}, {Watkins}, {Weaver}, {Whitmore},
  {Woillez}, {Zabalza}, \& {Astropy Contributors}}]{astropy_2018}
{Astropy Collaboration}, {Price-Whelan}, A.~M., {Sip{\H{o}}cz}, B.~M., {et~al.}
  2018, \aj, 156, 123, \dodoi{10.3847/1538-3881/aabc4f}

\bibitem[{{Barnes} \& {O'Brien}(2002)}]{Barnes+OBrien_2002}
{Barnes}, J.~W., \& {O'Brien}, D.~P. 2002, \apj, 575, 1087,
  \dodoi{10.1086/341477}

\bibitem[{{Barnes}(2017)}]{Barnes_2017}
{Barnes}, R. 2017, Celestial Mechanics and Dynamical Astronomy, 129, 509,
  \dodoi{10.1007/s10569-017-9783-7}

\bibitem[{{Barros} {et~al.}(2022){Barros}, {Akinsanmi}, {Bou{\'e}}, {Smith},
  {Laskar}, {Ulmer-Moll}, {Lillo-Box}, {Queloz}, {Cameron}, {Sousa},
  {Ehrenreich}, {Hooton}, {Bruno}, {Demory}, {Correia}, {Demangeon}, {Wilson},
  {Bonfanti}, {Hoyer}, {Alibert}, {Alonso}, {Escud{\'e}}, {Barbato},
  {B{\'a}rczy}, {Barrado}, {Baumjohann}, {Beck}, {Beck}, {Benz}, {Bergomi},
  {Billot}, {Bonfils}, {Bouchy}, {Brandeker}, {Broeg}, {Cabrera}, {Cessa},
  {Charnoz}, {Damme}, {Davies}, {Deleuil}, {Deline}, {Delrez}, {Erikson},
  {Fortier}, {Fossati}, {Fridlund}, {Gandolfi}, {Mu{\~n}oz}, {Gillon},
  {G{\"u}del}, {Isaak}, {Heng}, {Kiss}, {des Etangs}, {Lendl}, {Lovis},
  {Magrin}, {Nascimbeni}, {Maxted}, {Olofsson}, {Ottensamer}, {Pagano},
  {Pall{\'e}}, {Parviainen}, {Peter}, {Piotto}, {Pollacco}, {Ragazzoni},
  {Rando}, {Rauer}, {Ribas}, {Santos}, {Scandariato}, {S{\'e}gransan}, {Simon},
  {Steller}, {Szab{\'o}}, {Thomas}, {Udry}, {Ulmer}, {Van Grootel}, \&
  {Walton}}]{Barros+2022A&A}
{Barros}, S.~C.~C., {Akinsanmi}, B., {Bou{\'e}}, G., {et~al.} 2022, \aap, 657,
  A52, \dodoi{10.1051/0004-6361/202142196}

\bibitem[{{Becker} {et~al.}(2019){Becker}, {Vanderburg}, {Rodriguez},
  {Omohundro}, {Adams}, {Stassun}, {Yao}, {Hartman}, {Pepper}, {Bakos},
  {Barentsen}, {Beatty}, {Bhatti}, {Chontos}, {Collier Cameron}, {Hellier},
  {Huber}, {James}, {Kuhn}, {Lund}, {Pollacco}, {Siverd}, {Stevens}, {Cardoso},
  \& {West}}]{Becker+2019}
{Becker}, J.~C., {Vanderburg}, A., {Rodriguez}, J.~E., {et~al.} 2019, \aj, 157,
  19, \dodoi{10.3847/1538-3881/aaf0a2}

\bibitem[{{Belkovski} {et~al.}(2022){Belkovski}, {Becker}, {Howe}, {Malsky}, \&
  {Batygin}}]{Belkovski+2022}
{Belkovski}, M., {Becker}, J., {Howe}, A., {Malsky}, I., \& {Batygin}, K. 2022,
  \aj, 163, 277, \dodoi{10.3847/1538-3881/ac6353}

\bibitem[{{Berardo} {et~al.}(2019){Berardo}, {Crossfield}, {Werner},
  {Petigura}, {Christiansen}, {Ciardi}, {Dressing}, {Fulton}, {Gorjian},
  {Greene}, {Hardegree-Ullman}, {Kane}, {Livingston}, {Morales}, \&
  {Schlieder}}]{Berardo+2019}
{Berardo}, D., {Crossfield}, I. J.~M., {Werner}, M., {et~al.} 2019, \aj, 157,
  185, \dodoi{10.3847/1538-3881/ab100c}

\bibitem[{{Borucki} {et~al.}(2010){Borucki}, {Koch}, {Basri}, {Batalha},
  {Brown}, {Caldwell}, {Caldwell}, {Christensen-Dalsgaard}, {Cochran},
  {DeVore}, {Dunham}, {Dupree}, {Gautier}, {Geary}, {Gilliland}, {Gould},
  {Howell}, {Jenkins}, {Kondo}, {Latham}, {Marcy}, {Meibom}, {Kjeldsen},
  {Lissauer}, {Monet}, {Morrison}, {Sasselov}, {Tarter}, {Boss}, {Brownlee},
  {Owen}, {Buzasi}, {Charbonneau}, {Doyle}, {Fortney}, {Ford}, {Holman},
  {Seager}, {Steffen}, {Welsh}, {Rowe}, {Anderson}, {Buchhave}, {Ciardi},
  {Walkowicz}, {Sherry}, {Horch}, {Isaacson}, {Everett}, {Fischer}, {Torres},
  {Johnson}, {Endl}, {MacQueen}, {Bryson}, {Dotson}, {Haas}, {Kolodziejczak},
  {Van Cleve}, {Chandrasekaran}, {Twicken}, {Quintana}, {Clarke}, {Allen},
  {Li}, {Wu}, {Tenenbaum}, {Verner}, {Bruhweiler}, {Barnes}, \&
  {Prsa}}]{Borucki+2010}
{Borucki}, W.~J., {Koch}, D., {Basri}, G., {et~al.} 2010, Science, 327, 977,
  \dodoi{10.1126/science.1185402}

\bibitem[{{Bryant} {et~al.}(2021){Bryant}, {Bayliss}, {Santerne}, {Wheatley},
  {Nascimbeni}, {Ducrot}, {Burdanov}, {Acton}, {Alves}, {Anderson},
  {Armstrong}, {Awiphan}, {Cooke}, {Burleigh}, {Casewell}, {Delrez}, {Demory},
  {Eigm{\"u}ller}, {Fukui}, {Gan}, {Gill}, {Gillon}, {Goad}, {Tan},
  {G{\"u}nther}, {Hardee}, {Henderson}, {Jehin}, {Jenkins}, {Kosiarek},
  {Lendl}, {Moyano}, {Murray}, {Narita}, {Niraula}, {Odden}, {Palle},
  {Parviainen}, {Pedersen}, {Pozuelos}, {Rackham}, {Sebastian}, {Stockdale},
  {Tilbrook}, {Thompson}, {Triaud}, {Udry}, {Vines}, {West}, \& {de
  Wit}}]{Bryant+2021}
{Bryant}, E.~M., {Bayliss}, D., {Santerne}, A., {et~al.} 2021, \mnras, 504,
  L45, \dodoi{10.1093/mnrasl/slab037}

\bibitem[{{Buchner}(2016)}]{Buchner_2016}
{Buchner}, J. 2016, Statistics and Computing, 26, 383,
  \dodoi{10.1007/s11222-014-9512-y}

\bibitem[{{Buchner}(2019)}]{Buchner_2019}
---. 2019, \pasp, 131, 108005, \dodoi{10.1088/1538-3873/aae7fc}

\bibitem[{{Buchner}(2021)}]{Buchner_2021}
---. 2021, The Journal of Open Source Software, 6, 3001,
  \dodoi{10.21105/joss.03001}

\bibitem[{{Canup}(2010)}]{Canup+2010}
{Canup}, R.~M. 2010, \nat, 468, 943, \dodoi{10.1038/nature09661}

\bibitem[{{Cassese} \& {Kipping}(2022)}]{Cassese+2022}
{Cassese}, B., \& {Kipping}, D. 2022, \mnras, \dodoi{10.1093/mnras/stac2090}

\bibitem[{{Claret} \& {Gimenez}(1989)}]{Claret+1989AAS}
{Claret}, A., \& {Gimenez}, A. 1989, \aaps, 81, 37

\bibitem[{{Coulombe} {et~al.}(2023){Coulombe}, {Benneke}, {Challener},
  {Piette}, {Wiser}, {Mansfield}, {MacDonald}, {Beltz}, {Feinstein}, {Radica},
  {Savel}, {Dos Santos}, {Bean}, {Parmentier}, {Wong}, {Rauscher}, {Komacek},
  {Kempton}, {Tan}, {Hammond}, {Lewis}, {Line}, {Lee}, {Shivkumar},
  {Crossfield}, {Nixon}, {Rackham}, {Wakeford}, {Welbanks}, {Zhang}, {Batalha},
  {Berta-Thompson}, {Changeat}, {D{\'e}sert}, {Espinoza}, {Goyal},
  {Harrington}, {Knutson}, {Kreidberg}, {L{\'o}pez-Morales}, {Shporer}, {Sing},
  {Stevenson}, {Aggarwal}, {Ahrer}, {Alam}, {Bell}, {Blecic}, {Caceres},
  {Carter}, {Casewell}, {Crouzet}, {Cubillos}, {Decin}, {Fortney}, {Gibson},
  {Heng}, {Henning}, {Iro}, {Kendrew}, {Lagage}, {Leconte}, {Lendl},
  {Lothringer}, {Mancini}, {Mikal-Evans}, {Molaverdikhani}, {Nikolov}, {Ohno},
  {Palle}, {Piaulet}, {Redfield}, {Roy}, {Tsai}, {Venot}, \&
  {Wheatley}}]{Coulombe+2023arXiv}
{Coulombe}, L.-P., {Benneke}, B., {Challener}, R., {et~al.} 2023, arXiv
  e-prints, arXiv:2301.08192, \dodoi{10.48550/arXiv.2301.08192}

\bibitem[{{Crida} \& {Charnoz}(2012)}]{Crida+Sharnoz2012}
{Crida}, A., \& {Charnoz}, S. 2012, Science, 338, 1196,
  \dodoi{10.1126/science.1226477}

\bibitem[{{Dalba} {et~al.}(2015){Dalba}, {Muirhead}, {Fortney}, {Hedman},
  {Nicholson}, \& {Veyette}}]{Dalba+2015}
{Dalba}, P.~A., {Muirhead}, P.~S., {Fortney}, J.~J., {et~al.} 2015, \apj, 814,
  154, \dodoi{10.1088/0004-637X/814/2/154}

\bibitem[{{Domingos} {et~al.}(2006){Domingos}, {Winter}, \&
  {Yokoyama}}]{Domingos+2006}
{Domingos}, R.~C., {Winter}, O.~C., \& {Yokoyama}, T. 2006, \mnras, 373, 1227,
  \dodoi{10.1111/j.1365-2966.2006.11104.x}

\bibitem[{{Edwards} {et~al.}(2022){Edwards}, {Changeat}, {Tsiaras}, {Hou Yip},
  {Al-Refaie}, {Anisman}, {Bieger}, {Gressier}, {Shibata}, {Skaf}, {Bouwman},
  {Y-K. Cho}, {Ikoma}, {Venot}, {Waldmann}, {Lagage}, \&
  {Tinetti}}]{Edwards+2022arXiv}
{Edwards}, B., {Changeat}, Q., {Tsiaras}, A., {et~al.} 2022, arXiv e-prints,
  arXiv:2211.00649, \dodoi{10.48550/arXiv.2211.00649}

\bibitem[{{Forgan}(2017)}]{Forgan_2017}
{Forgan}, D.~H. 2017, \mnras, 470, 416, \dodoi{10.1093/mnras/stx1217}

\bibitem[{{Gao} \& {Zhang}(2020)}]{Gao+zhang_2020}
{Gao}, P., \& {Zhang}, X. 2020, \apj, 890, 93, \dodoi{10.3847/1538-4357/ab6a9b}

\bibitem[{{Greenberg}(2009)}]{Greenberg_2009}
{Greenberg}, R. 2009, \apjl, 698, L42, \dodoi{10.1088/0004-637X/698/1/L42}

\bibitem[{Harada(2023)}]{harada_astroQTpy_2023}
Harada, C.~K. 2023, astroQTpy, v0.2.0,  Zenodo, \dodoi{10.5281/zenodo.8187916}

\bibitem[{{Harris} {et~al.}(2020){Harris}, {Millman}, {van der Walt},
  {Gommers}, {Virtanen}, {Cournapeau}, {Wieser}, {Taylor}, {Berg}, {Smith},
  {Kern}, {Picus}, {Hoyer}, {van Kerkwijk}, {Brett}, {Haldane}, {del R{\'\i}o},
  {Wiebe}, {Peterson}, {G{\'e}rard-Marchant}, {Sheppard}, {Reddy}, {Weckesser},
  {Abbasi}, {Gohlke}, \& {Oliphant}}]{Harris+2020}
{Harris}, C.~R., {Millman}, K.~J., {van der Walt}, S.~J., {et~al.} 2020, \nat,
  585, 357, \dodoi{10.1038/s41586-020-2649-2}

\bibitem[{{Hellard} {et~al.}(2019){Hellard}, {Csizmadia}, {Padovan}, {Rauer},
  {Cabrera}, {Sohl}, {Spohn}, \& {Breuer}}]{Hellard+2019ApJ}
{Hellard}, H., {Csizmadia}, S., {Padovan}, S., {et~al.} 2019, \apj, 878, 119,
  \dodoi{10.3847/1538-4357/ab2048}

\bibitem[{{Heller}(2014)}]{Heller+2014_ose}
{Heller}, R. 2014, \apj, 787, 14, \dodoi{10.1088/0004-637X/787/1/14}

\bibitem[{{Heller} \& {Barnes}(2013)}]{Heller+Barnes_2013}
{Heller}, R., \& {Barnes}, R. 2013, Astrobiology, 13, 18,
  \dodoi{10.1089/ast.2012.0859}

\bibitem[{{Heller} {et~al.}(2016){Heller}, {Hippke}, \&
  {Jackson}}]{Heller+2016}
{Heller}, R., {Hippke}, M., \& {Jackson}, B. 2016, \apj, 820, 88,
  \dodoi{10.3847/0004-637X/820/2/88}

\bibitem[{{Heller} {et~al.}(2011){Heller}, {Leconte}, \&
  {Barnes}}]{Heller+2011A&A}
{Heller}, R., {Leconte}, J., \& {Barnes}, R. 2011, \aap, 528, A27,
  \dodoi{10.1051/0004-6361/201015809}

\bibitem[{{Heller} {et~al.}(2014){Heller}, {Williams}, {Kipping}, {Limbach},
  {Turner}, {Greenberg}, {Sasaki}, {Bolmont}, {Grasset}, {Lewis}, {Barnes}, \&
  {Zuluaga}}]{Heller+2014_review}
{Heller}, R., {Williams}, D., {Kipping}, D., {et~al.} 2014, Astrobiology, 14,
  798, \dodoi{10.1089/ast.2014.1147}

\bibitem[{{Hippke}(2015)}]{Hippke_2015}
{Hippke}, M. 2015, \apj, 806, 51, \dodoi{10.1088/0004-637X/806/1/51}

\bibitem[{{Hippke} \& {Heller}(2022)}]{Hippke+2022}
{Hippke}, M., \& {Heller}, R. 2022, \aap, 662, A37,
  \dodoi{10.1051/0004-6361/202243129}

\bibitem[{{Howell} {et~al.}(2014){Howell}, {Sobeck}, {Haas}, {Still},
  {Barclay}, {Mullally}, {Troeltzsch}, {Aigrain}, {Bryson}, {Caldwell},
  {Chaplin}, {Cochran}, {Huber}, {Marcy}, {Miglio}, {Najita}, {Smith},
  {Twicken}, \& {Fortney}}]{Howell+2014}
{Howell}, S.~B., {Sobeck}, C., {Haas}, M., {et~al.} 2014, \pasp, 126, 398,
  \dodoi{10.1086/676406}

\bibitem[{{Hunter}(2007)}]{Hunter+2007}
{Hunter}, J.~D. 2007, Computing in Science and Engineering, 9, 90,
  \dodoi{10.1109/MCSE.2007.55}

\bibitem[{{Hut}(1981)}]{Hut_1981}
{Hut}, P. 1981, \aap, 99, 126

\bibitem[{{Idini} \& {Stevenson}(2021)}]{Idini+2021}
{Idini}, B., \& {Stevenson}, D.~J. 2021, \psj, 2, 69,
  \dodoi{10.3847/PSJ/abe715}

\bibitem[{{Iess} {et~al.}(2019){Iess}, {Militzer}, {Kaspi}, {Nicholson},
  {Durante}, {Racioppa}, {Anabtawi}, {Galanti}, {Hubbard}, {Mariani},
  {Tortora}, {Wahl}, \& {Zannoni}}]{Iess+2019}
{Iess}, L., {Militzer}, B., {Kaspi}, Y., {et~al.} 2019, Science, 364, aat2965,
  \dodoi{10.1126/science.aat2965}

\bibitem[{{Jagtap} {et~al.}(2021){Jagtap}, {Quarles}, \& {Cuntz}}]{Jagtap+2021}
{Jagtap}, O., {Quarles}, B., \& {Cuntz}, M. 2021, \pasa, 38, e059,
  \dodoi{10.1017/pasa.2021.52}

\bibitem[{{Kipping}(2021)}]{Kipping_2021MNRAS}
{Kipping}, D. 2021, \mnras, 500, 1851, \dodoi{10.1093/mnras/staa3398}

\bibitem[{{Kipping} {et~al.}(2022){Kipping}, {Bryson}, {Burke}, {Christiansen},
  {Hardegree-Ullman}, {Quarles}, {Hansen}, {Szul{\'a}gyi}, \&
  {Teachey}}]{Kipping+2022}
{Kipping}, D., {Bryson}, S., {Burke}, C., {et~al.} 2022, Nature Astronomy, 6,
  367, \dodoi{10.1038/s41550-021-01539-1}

\bibitem[{{Kipping}(2011)}]{Kipping+2011}
{Kipping}, D.~M. 2011, \mnras, 416, 689,
  \dodoi{10.1111/j.1365-2966.2011.19086.x}

\bibitem[{{Kipping}(2013)}]{Kipping_2013MNRAS_limbdarkening}
---. 2013, \mnras, 435, 2152, \dodoi{10.1093/mnras/stt1435}

\bibitem[{{Kipping} {et~al.}(2012){Kipping}, {Bakos}, {Buchhave},
  {Nesvorn{\'y}}, \& {Schmitt}}]{Kipping+2012}
{Kipping}, D.~M., {Bakos}, G.~{\'A}., {Buchhave}, L., {Nesvorn{\'y}}, D., \&
  {Schmitt}, A. 2012, \apj, 750, 115, \dodoi{10.1088/0004-637X/750/2/115}

\bibitem[{{Kipping} {et~al.}(2013{\natexlab{a}}){Kipping}, {Forgan}, {Hartman},
  {Nesvorn{\'y}}, {Bakos}, {Schmitt}, \& {Buchhave}}]{Kipping+2013_III}
{Kipping}, D.~M., {Forgan}, D., {Hartman}, J., {et~al.} 2013{\natexlab{a}},
  \apj, 777, 134, \dodoi{10.1088/0004-637X/777/2/134}

\bibitem[{{Kipping} {et~al.}(2013{\natexlab{b}}){Kipping}, {Hartman},
  {Buchhave}, {Schmitt}, {Bakos}, \& {Nesvorn{\'y}}}]{Kipping+2013_II}
{Kipping}, D.~M., {Hartman}, J., {Buchhave}, L.~A., {et~al.}
  2013{\natexlab{b}}, \apj, 770, 101, \dodoi{10.1088/0004-637X/770/2/101}

\bibitem[{{Kipping} {et~al.}(2014){Kipping}, {Nesvorn{\'y}}, {Buchhave},
  {Hartman}, {Bakos}, \& {Schmitt}}]{Kipping+2014}
{Kipping}, D.~M., {Nesvorn{\'y}}, D., {Buchhave}, L.~A., {et~al.} 2014, \apj,
  784, 28, \dodoi{10.1088/0004-637X/784/1/28}

\bibitem[{{Kipping} {et~al.}(2015){Kipping}, {Schmitt}, {Huang}, {Torres},
  {Nesvorn{\'y}}, {Buchhave}, {Hartman}, \& {Bakos}}]{Kipping+2015}
{Kipping}, D.~M., {Schmitt}, A.~R., {Huang}, X., {et~al.} 2015, \apj, 813, 14,
  \dodoi{10.1088/0004-637X/813/1/14}

\bibitem[{{Kreidberg} {et~al.}(2019){Kreidberg}, {Luger}, \&
  {Bedell}}]{Kreidberg+2019}
{Kreidberg}, L., {Luger}, R., \& {Bedell}, M. 2019, \apjl, 877, L15,
  \dodoi{10.3847/2041-8213/ab20c8}

\bibitem[{{Lainey} {et~al.}(2017){Lainey}, {Jacobson}, {Tajeddine}, {Cooper},
  {Murray}, {Robert}, {Tobie}, {Guillot}, {Mathis}, {Remus}, {Desmars},
  {Arlot}, {De Cuyper}, {Dehant}, {Pascu}, {Thuillot}, {Le Poncin-Lafitte}, \&
  {Zahn}}]{Lainey+2017}
{Lainey}, V., {Jacobson}, R.~A., {Tajeddine}, R., {et~al.} 2017, \icarus, 281,
  286, \dodoi{10.1016/j.icarus.2016.07.014}

\bibitem[{{Laskar} {et~al.}(1993){Laskar}, {Joutel}, \&
  {Robutel}}]{Laskar+1993}
{Laskar}, J., {Joutel}, F., \& {Robutel}, P. 1993, \nat, 361, 615,
  \dodoi{10.1038/361615a0}

\bibitem[{{Leconte} {et~al.}(2010){Leconte}, {Chabrier}, {Baraffe}, \&
  {Levrard}}]{Leconte+2010A&A}
{Leconte}, J., {Chabrier}, G., {Baraffe}, I., \& {Levrard}, B. 2010, \aap, 516,
  A64, \dodoi{10.1051/0004-6361/201014337}

\bibitem[{{Limbach} {et~al.}(2022){Limbach}, {Soares-Furtado}, {Vanderburg},
  {Best}, {Cody}, {D'Onghia}, {Heller}, {Hensley}, {Kounkel}, {Kraus}, {Mann},
  {Robberto}, {Rosen}, {Townsend}, \& {Vos}}]{Limbach+2022}
{Limbach}, M.~A., {Soares-Furtado}, M., {Vanderburg}, A., {et~al.} 2022, arXiv
  e-prints, arXiv:2209.12916.
\newblock \doarXiv{2209.12916}

\bibitem[{{Lissauer} {et~al.}(2012){Lissauer}, {Barnes}, \&
  {Chambers}}]{Lissauer+2012}
{Lissauer}, J.~J., {Barnes}, J.~W., \& {Chambers}, J.~E. 2012, \icarus, 217,
  77, \dodoi{10.1016/j.icarus.2011.10.013}

\bibitem[{{Lustig-Yaeger} {et~al.}(2023){Lustig-Yaeger}, {Fu}, {May}, {Ortiz
  Ceballos}, {Moran}, {Peacock}, {Stevenson}, {L{\'o}pez-Morales}, {MacDonald},
  {Mayorga}, {Sing}, {Sotzen}, {Valenti}, {Adams}, {Alam}, {Batalha},
  {Bennett}, {Gonzalez-Quiles}, {Kirk}, {Kruse}, {Lothringer}, {Rustamkulov},
  \& {Wakeford}}]{Lustig-Yaeger+2023arXiv}
{Lustig-Yaeger}, J., {Fu}, G., {May}, E.~M., {et~al.} 2023, arXiv e-prints,
  arXiv:2301.04191, \dodoi{10.48550/arXiv.2301.04191}

\bibitem[{{Magic} {et~al.}(2015){Magic}, {Chiavassa}, {Collet}, \&
  {Asplund}}]{Magic+2015}
{Magic}, Z., {Chiavassa}, A., {Collet}, R., \& {Asplund}, M. 2015, \aap, 573,
  A90, \dodoi{10.1051/0004-6361/201423804}

\bibitem[{{Ment} \& {Charbonneau}(2023)}]{Ment+2023}
{Ment}, K., \& {Charbonneau}, D. 2023, \aj, 165, 265,
  \dodoi{10.3847/1538-3881/acd175}

\bibitem[{{Mignard}(1979)}]{Mignard_1979}
{Mignard}, F. 1979, Moon and Planets, 20, 301, \dodoi{10.1007/BF00907581}

\bibitem[{{Nakajima} {et~al.}(2020){Nakajima}, {Ida}, \&
  {Ishigaki}}]{Nakajima+2020}
{Nakajima}, A., {Ida}, S., \& {Ishigaki}, Y. 2020, \aap, 640, L15,
  \dodoi{10.1051/0004-6361/202038743}

\bibitem[{{Narang} {et~al.}(2022){Narang}, {Oza}, {Hakim}, {Puravankara},
  {Banyal}, \& {Thorngren}}]{Narang+2022arXiv}
{Narang}, M., {Oza}, A.~V., {Hakim}, K., {et~al.} 2022, arXiv e-prints,
  arXiv:2210.13298.
\newblock \doarXiv{2210.13298}

\bibitem[{{Ni}(2018)}]{Ni_2018}
{Ni}, D. 2018, \aap, 613, A32, \dodoi{10.1051/0004-6361/201732183}

\bibitem[{{Noyola} {et~al.}(2014){Noyola}, {Satyal}, \&
  {Musielak}}]{Noyola+2014}
{Noyola}, J.~P., {Satyal}, S., \& {Musielak}, Z.~E. 2014, \apj, 791, 25,
  \dodoi{10.1088/0004-637X/791/1/25}

\bibitem[{{Noyola} {et~al.}(2016){Noyola}, {Satyal}, \&
  {Musielak}}]{Noyola+2016}
---. 2016, \apj, 821, 97, \dodoi{10.3847/0004-637X/821/2/97}

\bibitem[{{Ohno} \& {Fortney}(2022)}]{Ohno+2022}
{Ohno}, K., \& {Fortney}, J.~J. 2022, \apj, 930, 50,
  \dodoi{10.3847/1538-4357/ac6029}

\bibitem[{{Ohno} \& {Tanaka}(2021)}]{Ohno+2021}
{Ohno}, K., \& {Tanaka}, Y.~A. 2021, \apj, 920, 124,
  \dodoi{10.3847/1538-4357/ac1516}

\bibitem[{{Ohno} {et~al.}(2022){Ohno}, {Thao}, {Mann}, \& {Fortney}}]{Ohno+22b}
{Ohno}, K., {Thao}, P.~C., {Mann}, A.~W., \& {Fortney}, J.~J. 2022, \apjl, 940,
  L30, \dodoi{10.3847/2041-8213/ac9f3f}

\bibitem[{{Owen} \& {Wu}(2016)}]{Owen+2016ApJ}
{Owen}, J.~E., \& {Wu}, Y. 2016, \apj, 817, 107,
  \dodoi{10.3847/0004-637X/817/2/107}

\bibitem[{{Payne} {et~al.}(2013){Payne}, {Deck}, {Holman}, \&
  {Perets}}]{Payne+2013}
{Payne}, M.~J., {Deck}, K.~M., {Holman}, M.~J., \& {Perets}, H.~B. 2013, \apjl,
  775, L44, \dodoi{10.1088/2041-8205/775/2/L44}

\bibitem[{{Perez} \& {Granger}(2007)}]{Perez+2007}
{Perez}, F., \& {Granger}, B.~E. 2007, Computing in Science and Engineering, 9,
  21, \dodoi{10.1109/MCSE.2007.53}

\bibitem[{{Peters} \& {Turner}(2013)}]{Peters+Turner_2013}
{Peters}, M.~A., \& {Turner}, E.~L. 2013, \apj, 769, 98,
  \dodoi{10.1088/0004-637X/769/2/98}

\bibitem[{{Piro}(2018{\natexlab{a}})}]{Piro+2018}
{Piro}, A.~L. 2018{\natexlab{a}}, \aj, 156, 54,
  \dodoi{10.3847/1538-3881/aaca38}

\bibitem[{{Piro}(2018{\natexlab{b}})}]{Piro_2018_rings}
---. 2018{\natexlab{b}}, \aj, 156, 80, \dodoi{10.3847/1538-3881/aad04a}

\bibitem[{{Piro} \& {Vissapragada}(2020)}]{Piro+Vissapragada_2020}
{Piro}, A.~L., \& {Vissapragada}, S. 2020, \aj, 159, 131,
  \dodoi{10.3847/1538-3881/ab7192}

\bibitem[{{Quarles} {et~al.}(2021){Quarles}, {Eggl}, {Rosario-Franco}, \&
  {Li}}]{Quarles+2021AJ}
{Quarles}, B., {Eggl}, S., {Rosario-Franco}, M., \& {Li}, G. 2021, \aj, 162,
  58, \dodoi{10.3847/1538-3881/ac042a}

\bibitem[{{Quarles} {et~al.}(2020){Quarles}, {Li}, \&
  {Rosario-Franco}}]{Quarles+2020}
{Quarles}, B., {Li}, G., \& {Rosario-Franco}, M. 2020, \apjl, 902, L20,
  \dodoi{10.3847/2041-8213/abba36}

\bibitem[{{Rackham} {et~al.}(2018){Rackham}, {Apai}, \&
  {Giampapa}}]{Rackham+2018}
{Rackham}, B.~V., {Apai}, D., \& {Giampapa}, M.~S. 2018, \apj, 853, 122,
  \dodoi{10.3847/1538-4357/aaa08c}

\bibitem[{{Rackham} {et~al.}(2019){Rackham}, {Apai}, \&
  {Giampapa}}]{Rackham+2019}
---. 2019, \aj, 157, 96, \dodoi{10.3847/1538-3881/aaf892}

\bibitem[{{Rauer} {et~al.}(2014){Rauer}, {Catala}, {Aerts}, {Appourchaux},
  {Benz}, {Brandeker}, {Christensen-Dalsgaard}, {Deleuil}, {Gizon}, {Goupil},
  {G{\"u}del}, {Janot-Pacheco}, {Mas-Hesse}, {Pagano}, {Piotto}, {Pollacco},
  {Santos}, {Smith}, {Su{\'a}rez}, {Szab{\'o}}, {Udry}, {Adibekyan}, {Alibert},
  {Almenara}, {Amaro-Seoane}, {Eiff}, {Asplund}, {Antonello}, {Barnes},
  {Baudin}, {Belkacem}, {Bergemann}, {Bihain}, {Birch}, {Bonfils}, {Boisse},
  {Bonomo}, {Borsa}, {Brand{\~a}o}, {Brocato}, {Brun}, {Burleigh}, {Burston},
  {Cabrera}, {Cassisi}, {Chaplin}, {Charpinet}, {Chiappini}, {Church},
  {Csizmadia}, {Cunha}, {Damasso}, {Davies}, {Deeg}, {D{\'\i}az}, {Dreizler},
  {Dreyer}, {Eggenberger}, {Ehrenreich}, {Eigm{\"u}ller}, {Erikson}, {Farmer},
  {Feltzing}, {de Oliveira Fialho}, {Figueira}, {Forveille}, {Fridlund},
  {Garc{\'\i}a}, {Giommi}, {Giuffrida}, {Godolt}, {Gomes da Silva}, {Granzer},
  {Grenfell}, {Grotsch-Noels}, {G{\"u}nther}, {Haswell}, {Hatzes},
  {H{\'e}brard}, {Hekker}, {Helled}, {Heng}, {Jenkins}, {Johansen},
  {Khodachenko}, {Kislyakova}, {Kley}, {Kolb}, {Krivova}, {Kupka}, {Lammer},
  {Lanza}, {Lebreton}, {Magrin}, {Marcos-Arenal}, {Marrese}, {Marques},
  {Martins}, {Mathis}, {Mathur}, {Messina}, {Miglio}, {Montalban}, {Montalto},
  {Monteiro}, {Moradi}, {Moravveji}, {Mordasini}, {Morel}, {Mortier},
  {Nascimbeni}, {Nelson}, {Nielsen}, {Noack}, {Norton}, {Ofir}, {Oshagh},
  {Ouazzani}, {P{\'a}pics}, {Parro}, {Petit}, {Plez}, {Poretti}, {Quirrenbach},
  {Ragazzoni}, {Raimondo}, {Rainer}, {Reese}, {Redmer}, {Reffert},
  {Rojas-Ayala}, {Roxburgh}, {Salmon}, {Santerne}, {Schneider}, {Schou},
  {Schuh}, {Schunker}, {Silva-Valio}, {Silvotti}, {Skillen}, {Snellen}, {Sohl},
  {Sousa}, {Sozzetti}, {Stello}, {Strassmeier}, {{\v{S}}vanda}, {Szab{\'o}},
  {Tkachenko}, {Valencia}, {Van Grootel}, {Vauclair}, {Ventura}, {Wagner},
  {Walton}, {Weingrill}, {Werner}, {Wheatley}, \& {Zwintz}}]{Rauer+2014}
{Rauer}, H., {Catala}, C., {Aerts}, C., {et~al.} 2014, Experimental Astronomy,
  38, 249, \dodoi{10.1007/s10686-014-9383-4}

\bibitem[{{Rein} \& {Liu}(2012)}]{rebound}
{Rein}, H., \& {Liu}, S.~F. 2012, \aap, 537, A128,
  \dodoi{10.1051/0004-6361/201118085}

\bibitem[{{Rein} \& {Spiegel}(2015)}]{reboundias15}
{Rein}, H., \& {Spiegel}, D.~S. 2015, \mnras, 446, 1424,
  \dodoi{10.1093/mnras/stu2164}

\bibitem[{{Ronnet} {et~al.}(2018){Ronnet}, {Mousis}, {Vernazza}, {Lunine}, \&
  {Crida}}]{Ronnet+2018}
{Ronnet}, T., {Mousis}, O., {Vernazza}, P., {Lunine}, J.~I., \& {Crida}, A.
  2018, \aj, 155, 224, \dodoi{10.3847/1538-3881/aabcc7}

\bibitem[{{Rosario-Franco} {et~al.}(2020){Rosario-Franco}, {Quarles},
  {Musielak}, \& {Cuntz}}]{Rosario-Franco+2020AJ}
{Rosario-Franco}, M., {Quarles}, B., {Musielak}, Z.~E., \& {Cuntz}, M. 2020,
  \aj, 159, 260, \dodoi{10.3847/1538-3881/ab89a7}

\bibitem[{{Saillenfest} {et~al.}(2023){Saillenfest}, {Sulis}, {Charpentier}, \&
  {Santerne}}]{Saillenfest+2023}
{Saillenfest}, M., {Sulis}, S., {Charpentier}, P., \& {Santerne}, A. 2023,
  \aap, 675, A174, \dodoi{10.1051/0004-6361/202346745}

\bibitem[{{Santerne} {et~al.}(2019){Santerne}, {Malavolta}, {Kosiarek}, {Dai},
  {Dressing}, {Dumusque}, {Hara}, {Lopez}, {Mortier}, {Vanderburg},
  {Adibekyan}, {Armstrong}, {Barrado}, {Barros}, {Bayliss}, {Berardo},
  {Boisse}, {Bonomo}, {Bouchy}, {Brown}, {Buchhave}, {Butler}, {Collier
  Cameron}, {Cosentino}, {Crane}, {Crossfield}, {Damasso}, {Deleuil}, {Delgado
  Mena}, {Demangeon}, {D{\'\i}az}, {Donati}, {Figueira}, {Fulton}, {Ghedina},
  {Harutyunyan}, {H{\'e}brard}, {Hirsch}, {Hojjatpanah}, {Howard}, {Isaacson},
  {Latham}, {Lillo-Box}, {L{\'o}pez-Morales}, {Lovis}, {Martinez Fiorenzano},
  {Molinari}, {Mousis}, {Moutou}, {Nava}, {Nielsen}, {Osborn}, {Petigura},
  {Phillips}, {Pollacco}, {Poretti}, {Rice}, {Santos}, {S{\'e}gransan},
  {Shectman}, {Sinukoff}, {Sousa}, {Sozzetti}, {Teske}, {Udry}, {Vigan},
  {Wang}, {Watson}, {Weiss}, {Wheatley}, \& {Winn}}]{Santerne+2019}
{Santerne}, A., {Malavolta}, L., {Kosiarek}, M.~R., {et~al.} 2019, arXiv
  e-prints, arXiv:1911.07355.
\newblock \doarXiv{1911.07355}

\bibitem[{{Sasaki} \& {Barnes}(2014)}]{Sasaki+Barnes_2014}
{Sasaki}, T., \& {Barnes}, J.~W. 2014, International Journal of Astrobiology,
  13, 324, \dodoi{10.1017/S1473550414000184}

\bibitem[{{Sasaki} {et~al.}(2012){Sasaki}, {Barnes}, \&
  {O'Brien}}]{Sasaki+2012}
{Sasaki}, T., {Barnes}, J.~W., \& {O'Brien}, D.~P. 2012, \apj, 754, 51,
  \dodoi{10.1088/0004-637X/754/1/51}

\bibitem[{{Sasaki} {et~al.}(2010){Sasaki}, {Stewart}, \& {Ida}}]{Sasaki+2010}
{Sasaki}, T., {Stewart}, G.~R., \& {Ida}, S. 2010, \apj, 714, 1052,
  \dodoi{10.1088/0004-637X/714/2/1052}

\bibitem[{{Showman} \& {Malhotra}(1999)}]{Showman+1999}
{Showman}, A.~P., \& {Malhotra}, R. 1999, Science, 296, 77

\bibitem[{{Spalding} {et~al.}(2016){Spalding}, {Batygin}, \&
  {Adams}}]{Spalding+2016}
{Spalding}, C., {Batygin}, K., \& {Adams}, F.~C. 2016, \apj, 817, 18,
  \dodoi{10.3847/0004-637X/817/1/18}

\bibitem[{{Sucerquia} {et~al.}(2019){Sucerquia}, {Alvarado-Montes}, {Zuluaga},
  {Cuello}, \& {Giuppone}}]{Sucerquia+2019MNRAS}
{Sucerquia}, M., {Alvarado-Montes}, J.~A., {Zuluaga}, J.~I., {Cuello}, N., \&
  {Giuppone}, C. 2019, \mnras, 489, 2313, \dodoi{10.1093/mnras/stz2110}

\bibitem[{{Sucerquia} {et~al.}(2020){Sucerquia}, {Ram{\'\i}rez},
  {Alvarado-Montes}, \& {Zuluaga}}]{Sucerquia+2020MNRAS}
{Sucerquia}, M., {Ram{\'\i}rez}, V., {Alvarado-Montes}, J.~A., \& {Zuluaga},
  J.~I. 2020, \mnras, 492, 3499, \dodoi{10.1093/mnras/stz3548}

\bibitem[{{Teachey} {et~al.}(2020){Teachey}, {Kipping}, {Burke}, {Angus}, \&
  {Howard}}]{Teachey+2020}
{Teachey}, A., {Kipping}, D., {Burke}, C.~J., {Angus}, R., \& {Howard}, A.~W.
  2020, \aj, 159, 142, \dodoi{10.3847/1538-3881/ab7001}

\bibitem[{{Teachey} \& {Kipping}(2018)}]{Teachey+2018_nature}
{Teachey}, A., \& {Kipping}, D.~M. 2018, Science Advances, 4, eaav1784,
  \dodoi{10.1126/sciadv.aav1784}

\bibitem[{{Teachey} {et~al.}(2018){Teachey}, {Kipping}, \&
  {Schmitt}}]{Teachey+2018}
{Teachey}, A., {Kipping}, D.~M., \& {Schmitt}, A.~R. 2018, \aj, 155, 36,
  \dodoi{10.3847/1538-3881/aa93f2}

\bibitem[{{The LUVOIR Team}(2019)}]{LUVOIR_2019arXiv}
{The LUVOIR Team}. 2019, arXiv e-prints, arXiv:1912.06219,
  \dodoi{10.48550/arXiv.1912.06219}

\bibitem[{{Tokadjian} \& {Piro}(2020)}]{Tokadjian+2020}
{Tokadjian}, A., \& {Piro}, A.~L. 2020, \aj, 160, 194,
  \dodoi{10.3847/1538-3881/abb29e}

\bibitem[{{van der Walt} {et~al.}(2011){van der Walt}, {Colbert}, \&
  {Varoquaux}}]{vanderWalt+2011}
{van der Walt}, S., {Colbert}, S.~C., \& {Varoquaux}, G. 2011, Computing in
  Science and Engineering, 13, 22, \dodoi{10.1109/MCSE.2011.37}

\bibitem[{{Vanderburg} {et~al.}(2018){Vanderburg}, {Rappaport}, \&
  {Mayo}}]{Vanderburg+2018}
{Vanderburg}, A., {Rappaport}, S.~A., \& {Mayo}, A.~W. 2018, \aj, 156, 184,
  \dodoi{10.3847/1538-3881/aae0fc}

\bibitem[{{Vanderburg} {et~al.}(2016{\natexlab{a}}){Vanderburg}, {Becker},
  {Kristiansen}, {Bieryla}, {Duev}, {Jensen-Clem}, {Morton}, {Latham}, {Adams},
  {Baranec}, {Berlind}, {Calkins}, {Esquerdo}, {Kulkarni}, {Law}, {Riddle},
  {Salama}, \& {Schmitt}}]{Vanderburg+2016_hip}
{Vanderburg}, A., {Becker}, J.~C., {Kristiansen}, M.~H., {et~al.}
  2016{\natexlab{a}}, \apjl, 827, L10, \dodoi{10.3847/2041-8205/827/1/L10}

\bibitem[{{Vanderburg} {et~al.}(2016{\natexlab{b}}){Vanderburg}, {Latham},
  {Buchhave}, {Bieryla}, {Berlind}, {Calkins}, {Esquerdo}, {Welsh}, \&
  {Johnson}}]{Vanderburg+2016_k2sff}
{Vanderburg}, A., {Latham}, D.~W., {Buchhave}, L.~A., {et~al.}
  2016{\natexlab{b}}, \apjs, 222, 14, \dodoi{10.3847/0067-0049/222/1/14}

\bibitem[{Wakeford \& Grant(2022)}]{hannah_wakeford_2022_6809899}
Wakeford, H., \& Grant, D. 2022, Exo-TiC/ExoTiC-LD: ExoTiC-LD v2.1 Zenodo
  Release, v2.1.0,  Zenodo, \dodoi{10.5281/zenodo.6809899}

\bibitem[{{Wang} \& {Dai}(2019)}]{Wang+Dai_2019ApJ}
{Wang}, L., \& {Dai}, F. 2019, \apjl, 873, L1, \dodoi{10.3847/2041-8213/ab0653}

\bibitem[{{Wheatley} {et~al.}(2018){Wheatley}, {West}, {Goad}, {Jenkins},
  {Pollacco}, {Queloz}, {Rauer}, {Udry}, {Watson}, {Chazelas}, {Eigm{\"u}ller},
  {Lambert}, {Genolet}, {McCormac}, {Walker}, {Armstrong}, {Bayliss}, {Bento},
  {Bouchy}, {Burleigh}, {Cabrera}, {Casewell}, {Chaushev}, {Chote},
  {Csizmadia}, {Erikson}, {Faedi}, {Foxell}, {G{\"a}nsicke}, {Gillen},
  {Grange}, {G{\"u}nther}, {Hodgkin}, {Jackman}, {Jord{\'a}n}, {Louden},
  {Metrailler}, {Moyano}, {Nielsen}, {Osborn}, {Poppenhaeger}, {Raddi},
  {Raynard}, {Smith}, {Soto}, \& {Titz-Weider}}]{Wheatley+2018MNRAS}
{Wheatley}, P.~J., {West}, R.~G., {Goad}, M.~R., {et~al.} 2018, \mnras, 475,
  4476, \dodoi{10.1093/mnras/stx2836}

\bibitem[{{Zhang} {et~al.}(2019){Zhang}, {Chachan}, {Kempton}, \&
  {Knutson}}]{zhang+2019PASP_platon}
{Zhang}, M., {Chachan}, Y., {Kempton}, E. M.~R., \& {Knutson}, H.~A. 2019,
  \pasp, 131, 034501, \dodoi{10.1088/1538-3873/aaf5ad}

\bibitem[{{Zhang} {et~al.}(2020){Zhang}, {Chachan}, {Kempton}, {Knutson}, \&
  {Chang}}]{Zhang+2020ApJ_platon}
{Zhang}, M., {Chachan}, Y., {Kempton}, E. M.~R., {Knutson}, H.~A., \& {Chang},
  W.~H. 2020, \apj, 899, 27, \dodoi{10.3847/1538-4357/aba1e6}

\bibitem[{{Zuluaga} {et~al.}(2015){Zuluaga}, {Kipping}, {Sucerquia}, \&
  {Alvarado}}]{Zuluaga+2015ApJ}
{Zuluaga}, J.~I., {Kipping}, D.~M., {Sucerquia}, M., \& {Alvarado}, J.~A. 2015,
  \apjl, 803, L14, \dodoi{10.1088/2041-8205/803/1/L14}

\end{thebibliography}
\bibliographystyle{aasjournal}

\end{document}